%% file: conf.tex
\documentclass[11pt]{article}
\usepackage{graphicx}

\newcommand{\BABARPubYear}    {06}

\newcommand{\BABARConfNumber} {40}
\newcommand{\SLACPubNumber} {12015}

\input babarsym

\def\KKKz      {\ensuremath{K^+ K^- K^0}\xspace}
\def\KKKs      {\ensuremath{K^+ K^- \KS}\xspace}
\def\KKKl      {\ensuremath{K^+ K^- \KL}\xspace}
\def\KKKspm      {\ensuremath{K^+ K^- {\KS}(\pip\pim)}\xspace}
\def\KKKszz      {\ensuremath{K^+ K^- {\KS}(\piz\piz)}\xspace}

\def\sPlot{\ensuremath{_s{\cal P}lot}\xspace}
\def\splot{\ensuremath{_s{\cal P}lot}\xspace}

\def\mKK       {\ensuremath{m_{\Kp\Km}}\xspace}

\def\betaeff   {\ensuremath{\beta_{\mathit{eff}}}\xspace}
\def\Acp   {\ensuremath{{A}_{\CP}}\xspace}
\def\cosH      {\ensuremath{\cos \theta_H}\xspace}

\def\spk{\ensuremath{S_{\phi K}}\xspace}
\def\cpk{\ensuremath{C_{\phi K}}\xspace}

\long\def\inst#1{\par\nobreak\kern 4pt\nobreak
    {\it #1}\par\vskip 10pt plus 3pt minus 3pt}

\begin{document}
{\pagestyle{empty}


\begin{flushright}
\babar-CONF-\BABARPubYear/\BABARConfNumber \\
SLAC-PUB-\SLACPubNumber \\
\end{flushright}

\par\vskip 3.75cm

\begin{center}
\Large \bf Measurement of {\boldmath$\CP$}-Violating Asymmetries in the {\boldmath $\Bz\to\Kp\Km\Kz$} Dalitz Plot
\end{center}
\bigskip

\begin{center}
\large The \babar\ Collaboration\\
\mbox{ }\\
\today
\end{center}
\bigskip \bigskip

\begin{center}
\large \bf Abstract
\end{center}
We present a preliminary measurement of \CP-violation parameters in the decay $\Bz \to \KKKz$,
using approximately 347 million $\BB$ events collected by the $\babar$ detector at SLAC. 
Reconstructing the neutral kaon as $\KS \to \pip\pim, \KS \to \piz\piz$, or
$\KL$, we analyze the Dalitz plot distribution and measure fractions
to intermediate states.  We extract \CP parameters from the
asymmetries in amplitudes and phases between $\Bz$ and $\Bzb$ decays
across the Dalitz plot.  For decays to $\phi\Kz$, we find
$\betaeff=0.06 \pm 0.16 \pm 0.05$, $\Acp=-0.18 \pm 0.20 \pm 0.10$,
where the first uncertainty is statistical and the second one is
systematic.  For decays to $f_0\Kz$, we find $\betaeff=0.18 \pm 0.19
\pm 0.04$, $\Acp=0.45 \pm 0.28 \pm 0.10$.  Combining all \KKKz
events and taking account of the different \CP eigenvalues of the individual Dalitz plot 
components, we find $\betaeff = 0.361 \pm 0.079 \pm 0.037,~
\Acp = -0.034 \pm 0.079 \pm 0.025$. The trigonometric reflection at
$\pi/2 - \betaeff$ is disfavored at $4.6\sigma$.  We also study angular
distributions in $\Bz\to\KKKs$ and $\Bp \to\phi\Kp$ decays and measure
the direct \CP asymmetry in $\Bp \to\phi\Kp$ decays, $\Acp=0.046
\pm 0.046 \pm 0.017$.

\vfill
\begin{center}

Submitted to the 33$^{\rm rd}$ International Conference on High-Energy Physics, ICHEP 06,\\
26 July---2 August 2006, Moscow, Russia.

\end{center}

\vspace{1.0cm}
\begin{center}
{\em Stanford Linear Accelerator Center, Stanford University, 
Stanford, CA 94309} \\ \vspace{0.1cm}\hrule\vspace{0.1cm}
Work supported in part by Department of Energy contract DE-AC03-76SF00515.
\end{center}

\newpage
} 


\input authors_ICHEP2006.tex

\section{INTRODUCTION}
\label{sec:Introduction}

We describe a \B-flavor tagged, time-dependent Dalitz plot analysis of the $\Bz \to
\KKKz$ decay~\cite{conjugate}, with the \Kz reconstructed as $\KS \to \pip \pim$, $\KS \to \piz \piz$,
or \KL.  In the Standard Model (SM), these decays are dominated by 
$\b \to s\bar{s}s$ gluonic penguin amplitudes, with a single weak
phase.  Contributions from $b\to u \bar{q}q$ tree amplitudes,
proportional to the Cabibbo-Kobayashi-Maskawa (CKM) matrix element
$V_{ub}$ with a \CP-violating weak phase $\gamma$~\cite{Eidelman:2004wy}, are small, but may
depend on the position in the Dalitz plot.  In $\Bz\to\phi(\Kp\Km)\Kz$
decays the modification of the \CP asymmetry due to the presence of
suppressed tree amplitudes is at $\cal
O$(0.01)~\cite{Beneke:2005pu,Buchalla:2005us}, while at higher $\Kp\Km$ masses a larger
contribution at $\cal O$(0.1) is possible~\cite{Cheng:2005ug}.
Therefore, to very good precision, we also expect the direct \CP
asymmetry for these decays to be small in the SM.  The \CP asymmetry
in $\Bz \to \KKKz$ decay arises from the interference
of decays and $\Bz \leftrightarrow \Bzb$ mixing, with a relative phase
of $2\beta$.  The Unitarity Triangle angle $\beta$ has been measured
in  $\Bz\to [c\bar{c}]\Kz$ decays to be  
$\sin2\beta=0.685 \pm 0.032$~\cite{Aubert:2004zt,Abe:2005bt}.
Current direct measurements favor the solution of  $\beta=0.37$
over $\beta=1.20$ at the 98.3\% C.L.~\cite{Krokovny:2006sv}.

The decay $\Bz \to \KKKz$ is one of the most promising processes with which to
search for physics beyond the SM.  Since  the leading  amplitudes
enter only at the one-loop level, additional contributions from heavy
non-SM particles may be of comparable size. If the amplitude from
heavy particles has a \CP-violating phase, the measured \CP-violation
parameters may differ from those expected in the SM.

Previous measurements of the \CP asymmetry in $\Bz \to \KKKz$ decays
have been performed separately around the $\phi$ mass, and for higher
$\Kp\Km$ masses, neglecting interference effects between intermediate
states~\cite{Aubert:2005ja}.  In this analysis, we extract the
\CP-violation parameters by taking into account different amplitudes
and phases across the \Bz and \Bzb Dalitz plots.

%
%

\section{EVENT RECONSTRUCTION}
\label{sec:selection}

The data used in this analysis were collected with the \babar\ detector
at the \pep2\ asymmetric-energy \B factory at SLAC. A total of 347
million \BB pairs were used.

The \babar\ detector is described in detail
elsewhere~\cite{ref:babar}.  Charged particle (track) momenta are
measured with a 5-layer double-sided silicon vertex tracker (SVT) and a
40-layer drift chamber (DCH) coaxial with a 1.5-T superconducting solenoidal
magnet.  Neutral cluster (photon) positions and energies are measured
with an electromagnetic calorimeter (EMC) consisting of 6580 CsI(Tl)
crystals.  Charged hadrons are identified with a detector of
internally reflected Cherenkov light (DIRC) and specific ionization
measurements (\dedx) in the tracking detectors (DCH, SVT).  Neutral hadrons that do not
interact in the EMC are identified with detectors, up to 15 layers
deep, in the flux return steel (IFR). 


We reconstruct $\Bz \to \KKKz$ decays by combining two oppositely
charged tracks with a $\KS\to\pip\pim$, $\KS\to\piz\piz$, or $\KL$
candidate. $\Bp \to \phi\Kp$ decays are reconstructed from three charged
tracks. The \Kp and \Km tracks must have at least 12 measured DCH
coordinates, a minimum transverse momentum of
0.1~\gevc, and must originate from  the nominal beam spot.  Tracks
are identified as kaons using a likelihood ratio
that combines \dedx measured in the SVT and DCH with the Cherenkov
angle and number of photons measured in the DIRC. For $\Kp\Km$ masses
near the $\phi$ mass, higher efficiency kaon identification is used, while for higher masses
higher purity criteria are chosen to reduce background.  

For all modes, the main background is from random combinations of
particles produced in events of the type $e^+e^-\to q\bar{q}~
(q=u,d,s,c)$ (continuum).  Additional background from decays of $B$
mesons to other final states, with and without charm particles, is
estimated from Monte Carlo simulation.

We use event-shape variables, computed in the center-of-mass (CM)
frame, to separate continuum events with a jet-like topology from the
more isotropic \B decays.  Continuum events are suppressed with a
requirement on the quantity $\cos (\theta_{\rm T})$, $\cos
(\theta_{\rm T}) < 0.9$, where $\theta_{\rm T}$ is the angle between
the thrust axis of the $B$ candidate's daughters and the thrust axis
formed from the other charged and neutral particles in the event. We
select events in the range . Further discrimination comes from the
Legendre moments $\mathcal{L}_{i=0,2} = \sum_j p_j L_{i}(\theta_{j})$,
where the sum is over all tracks and clusters not used to reconstruct
the \B meson; $L_i$ is the Legendre polynomial of order $i$, and
$\theta_{j}$ is the angle to the \B thrust axis.  Lastly, the
magnitude of the cosine of the angle of the
\B with respect to the collision axis $|\cos{\theta_{B}}|$, is also used.

\subsection{{\boldmath $\Bz \to \Kp\Km\KS$, $\KS\to\pip\pim$}}
\label{sec:kkkspm_selection}

For decays $\Bz \to \Kp\Km\KS$ and $\KS\to\pip\pim$, \KS candidates
are formed from oppositely charged tracks with an invariant mass within
$20~\mevcc$ of the \KS mass~\cite{Eidelman:2004wy}.  The \KS vertex is required to be
separated from the \Bz vertex by at
least $3\sigma$.  The angle $\alpha$ between the \KS momentum vector and the
vector connecting the \Bz and \KS vertices must satisfy $\cos\alpha >
0.999$.

\begin{figure}[ptb]
\center
\begin{tabular}{ll}
\includegraphics[height=5.0cm]{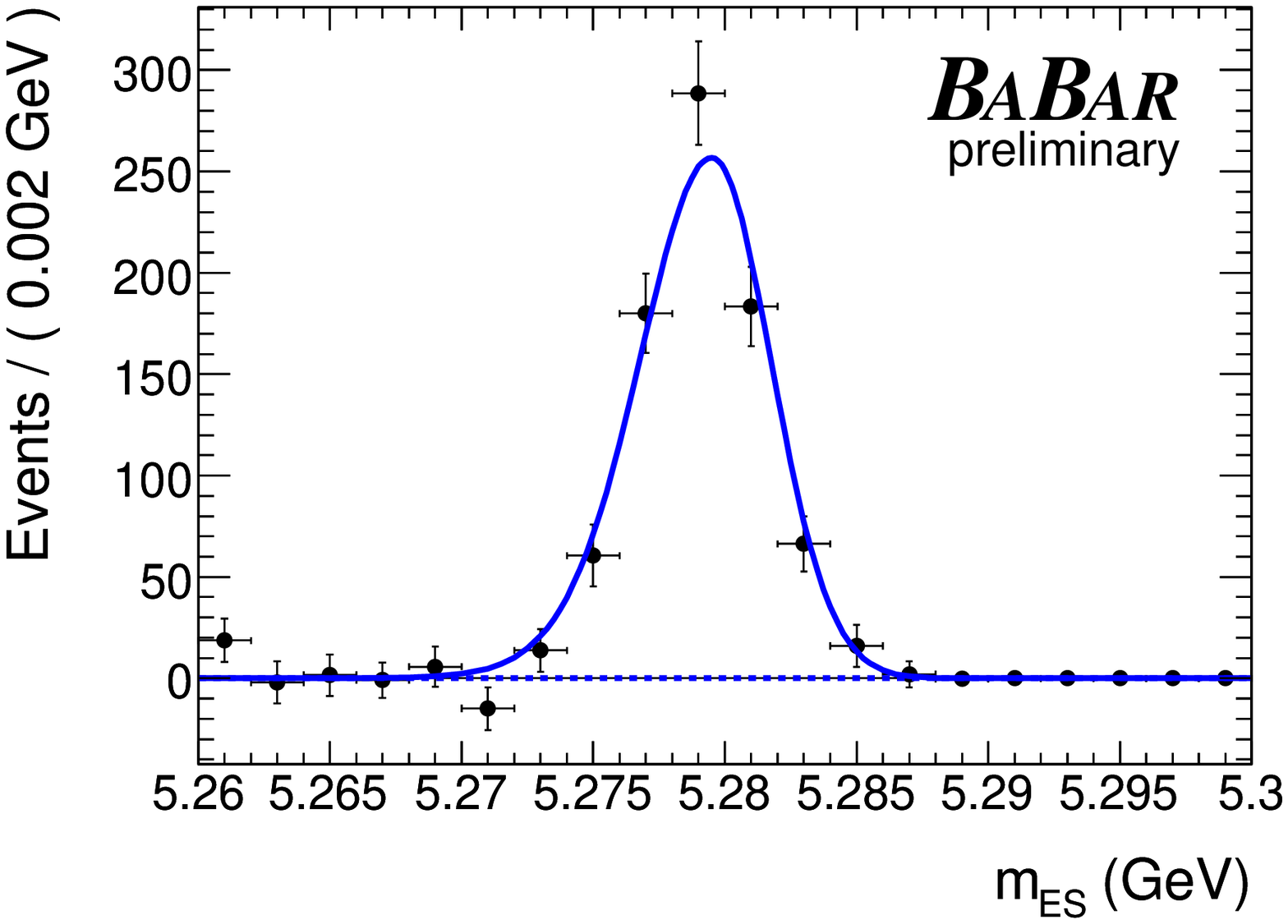} & \includegraphics[height=5.0cm]{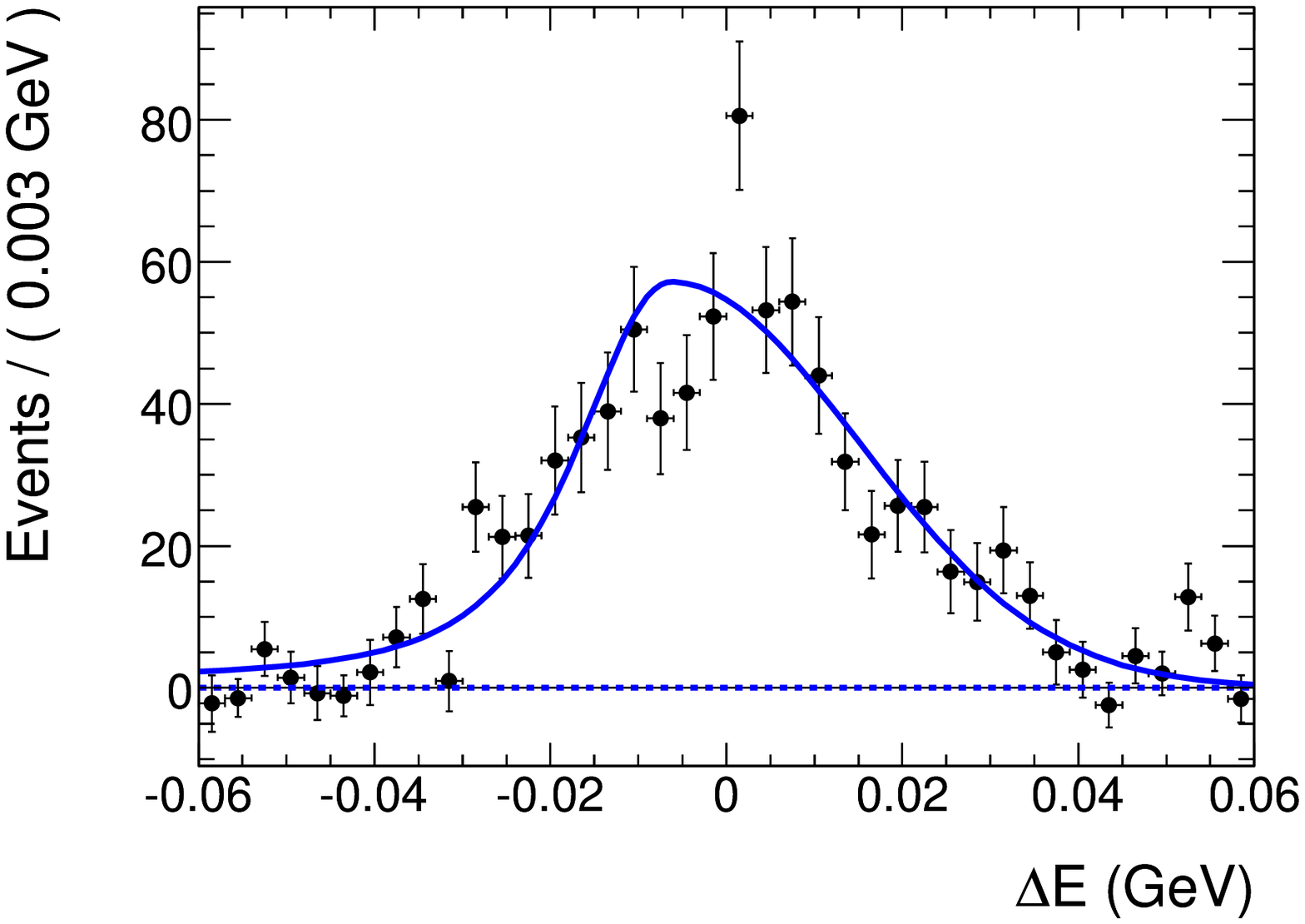} \\
\includegraphics[height=5.0cm]{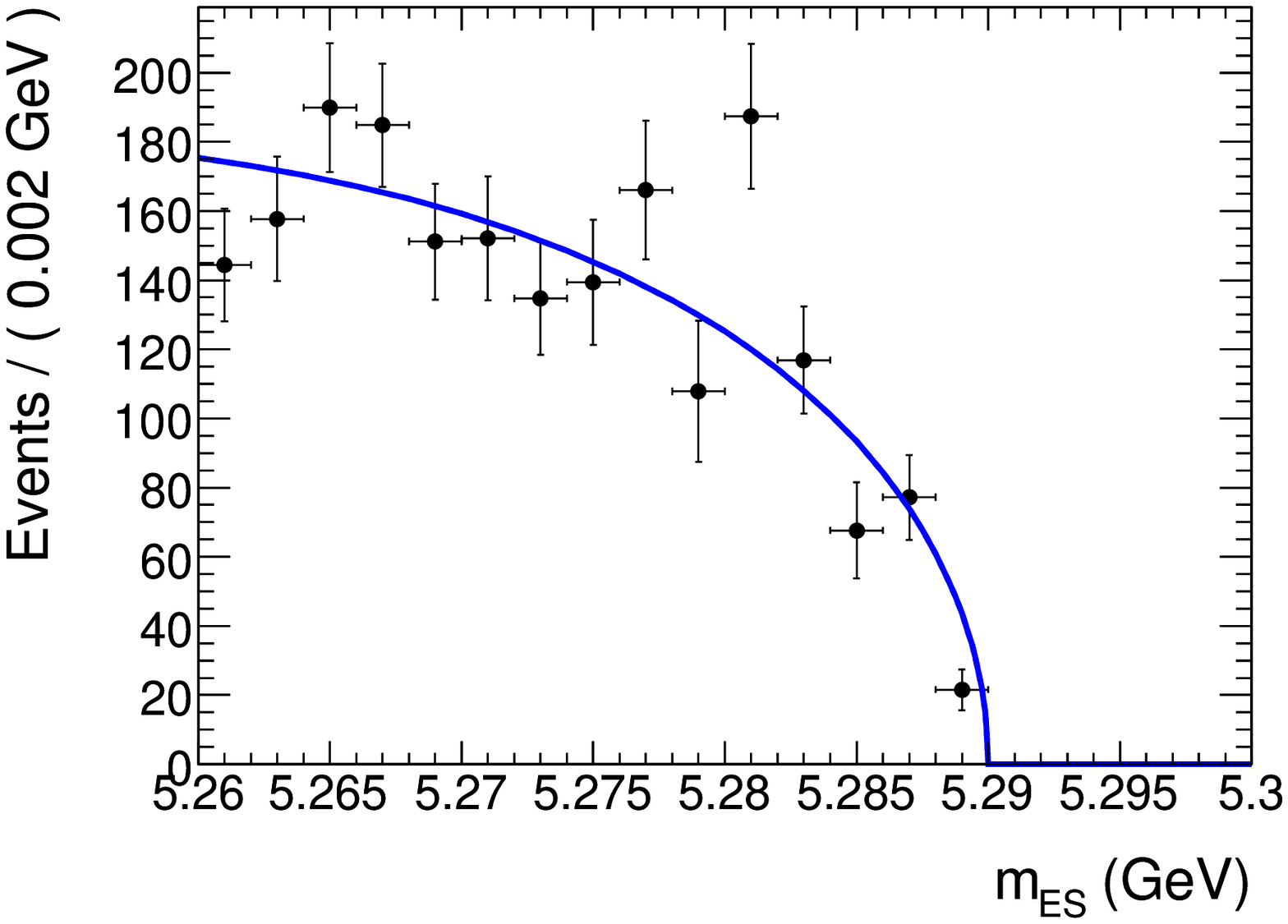} & \includegraphics[height=5.0cm]{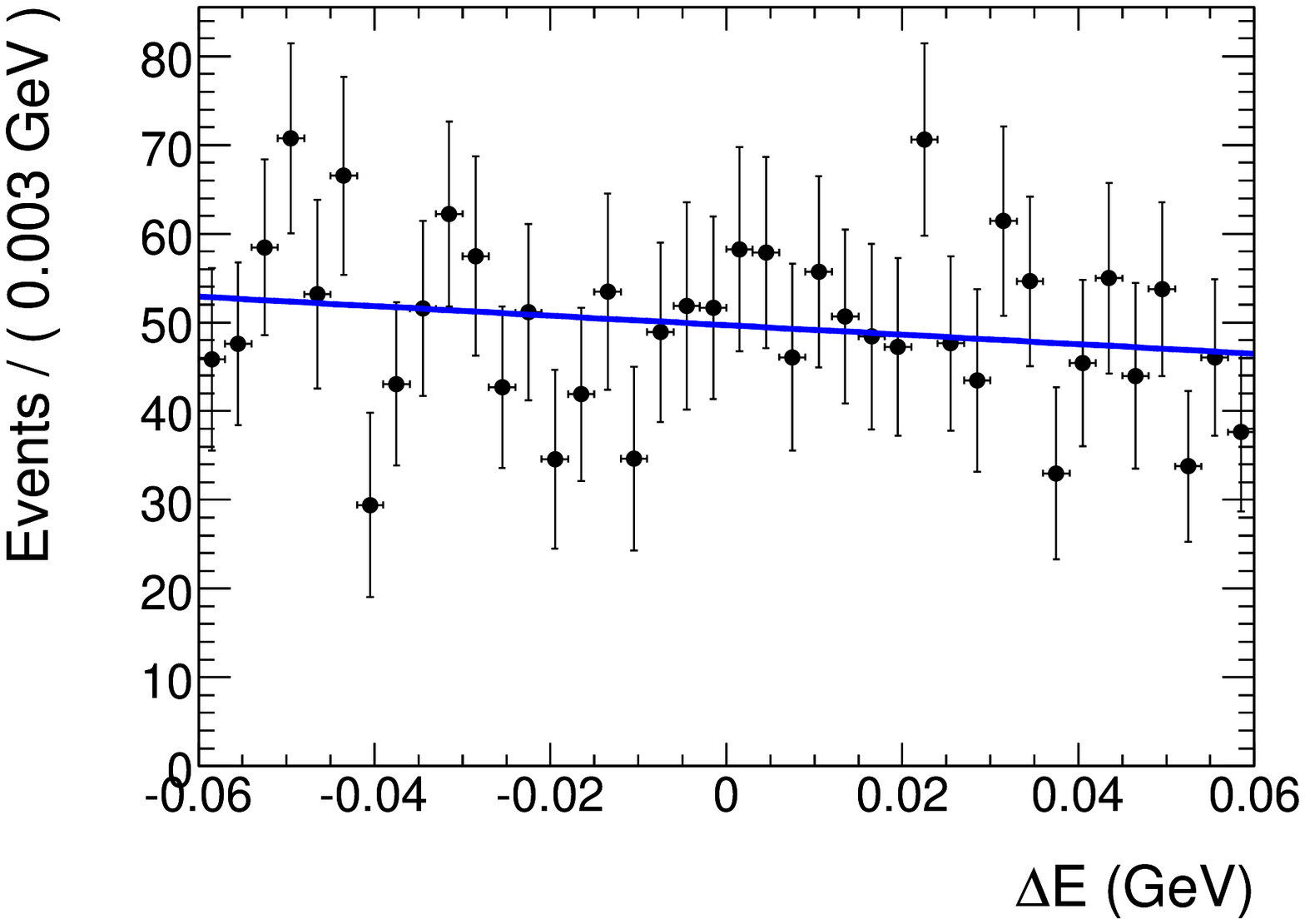} \\
\end{tabular}
\caption{Distributions of kinematic variables (left) \mes\ and (right) \DeltaE for the \KKKspm subsample: 
(top) signal, (bottom)  continuum background. The points are data events weighted with the \sPlot\ technique~\cite{Pivk:2004ty}, 
and the curves are the PDF shapes used in the ML fit (Sec. \ref{sec:Dalitz}).}
\label{fg::kkks+-_event_selection}
\end{figure}

\B candidates are identified using two kinematic variables that separate
signal from continuum background. These are the beam-energy-substituted mass
$\mes = \sqrt{ (s/2 + {\bf p}_{i}\cdot{\bf p}_{B})^{2}/E_{i}^{2}- {\bf
    p}^{2}_{B}}$, where $\sqrt{s}$ is the total \epem CM
(CM) energy, $(E_{i},{\bf p}_{i})$ is the four-momentum of the initial
\epem system and ${\bf p}_{B}$ is the \B candidate momentum, both
measured in the laboratory frame, and $\Delta E = E_{B} - \sqrt{s}/2$,
where $E_{B}$ is the \B candidate energy in the CM
frame. Distributions of these variables in data, for signal and
background events calculated using the \splot event-weighting
technique~\cite{Pivk:2004ty}, are shown in Fig.~\ref{fg::kkks+-_event_selection}.

\subsection{{\boldmath $\Bz \to \Kp\Km\KS$, $\KS\to\piz\piz$}}

For decays $\Bz \to \Kp\Km\KS$ and $\KS\to\piz\piz$, \KS candidates
are formed from two $\piz\to\gamma\gamma$ candidates.  Each of the
four photons must have $E_{\gamma} > 0.05 \gev$ and have a transverse
shower shape loosely consistent with an electromagnetic
shower. Additionally, we require each \piz candidate to satisfy $0.100
< m_{\gamma\gamma} < 0.155 \gevcc$. The resulting $\KS\to\piz\piz$
mass is required to satisfy $0.4776 < m_{\piz\piz} < 0.5276~\gevcc$. A \KS
mass constraint is then applied for the reconstruction of the \Bz
candidate.

\begin{figure}[ptb]
\center
\begin{tabular}{ll}
\includegraphics[height=5.0cm]{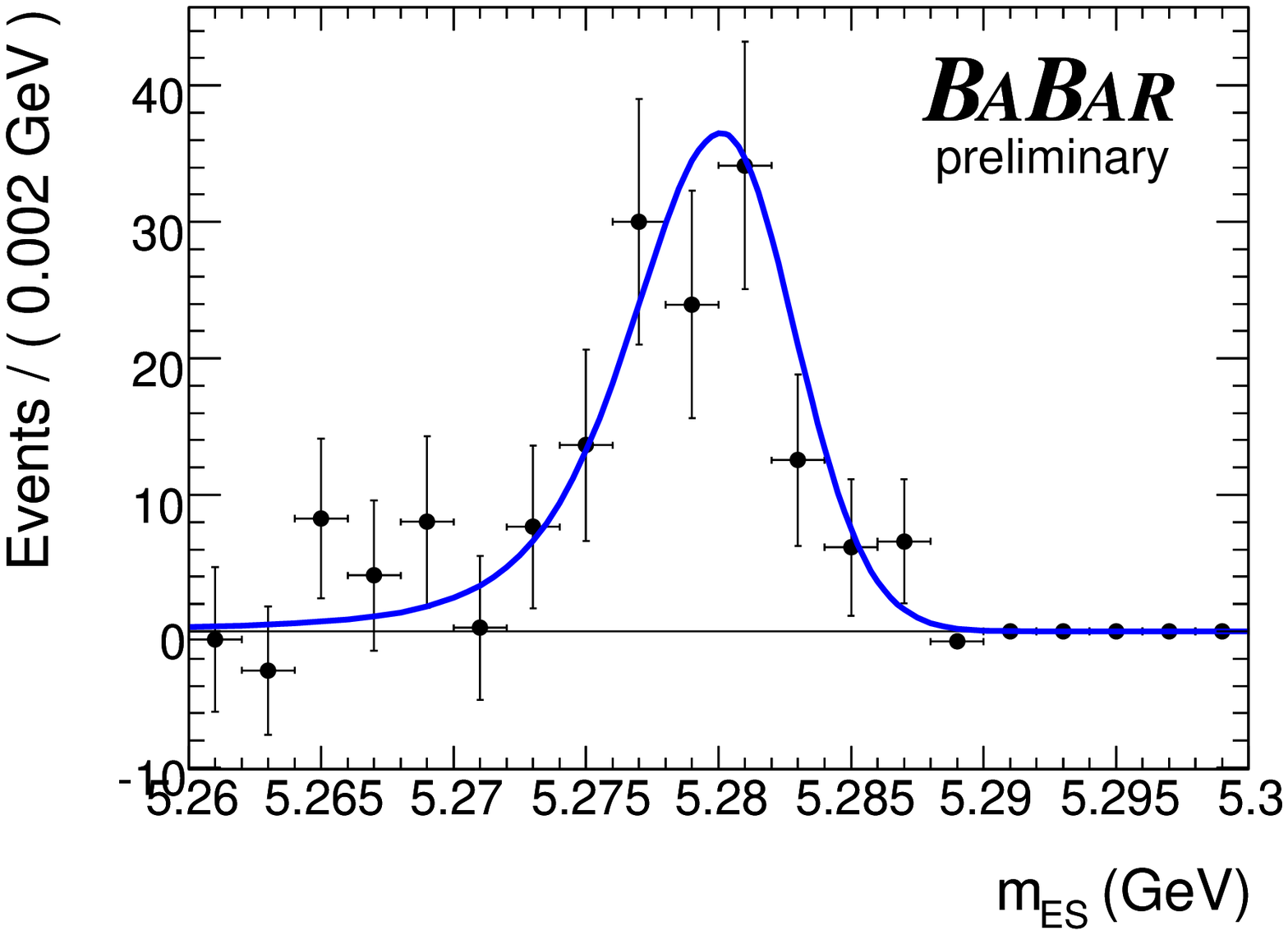} & \includegraphics[height=5.0cm]{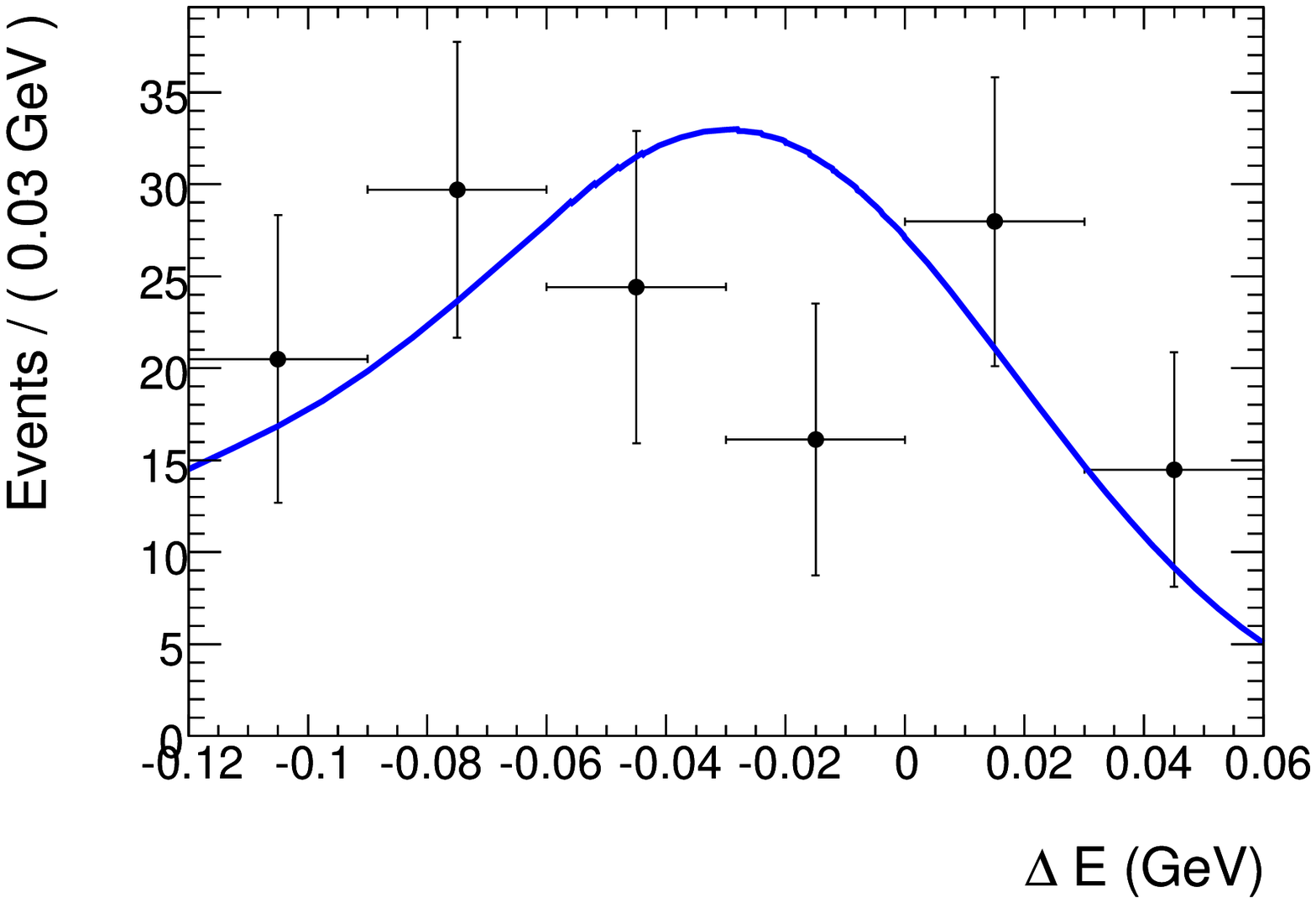} \\
\includegraphics[height=5.0cm]{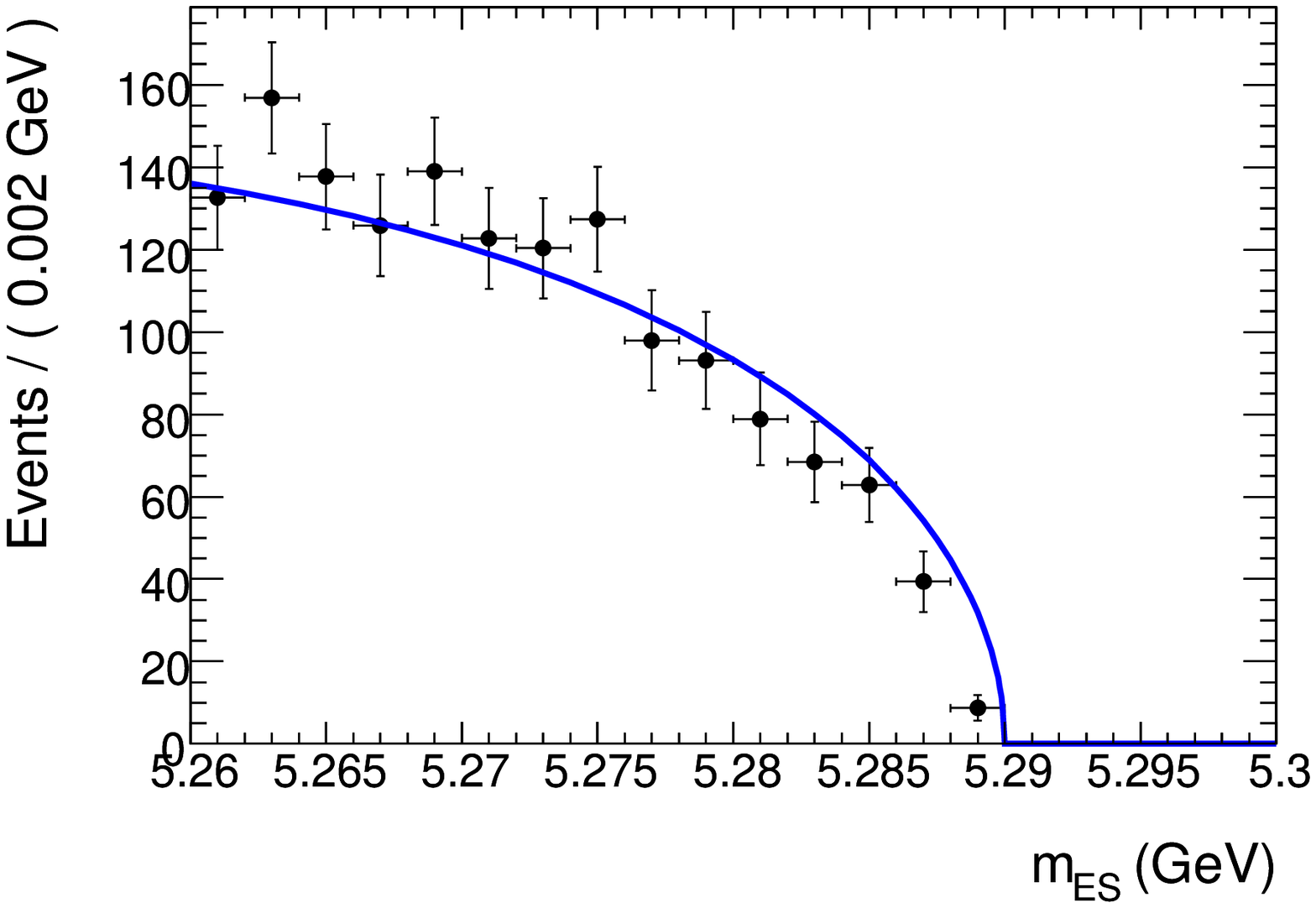} & \includegraphics[height=5.0cm]{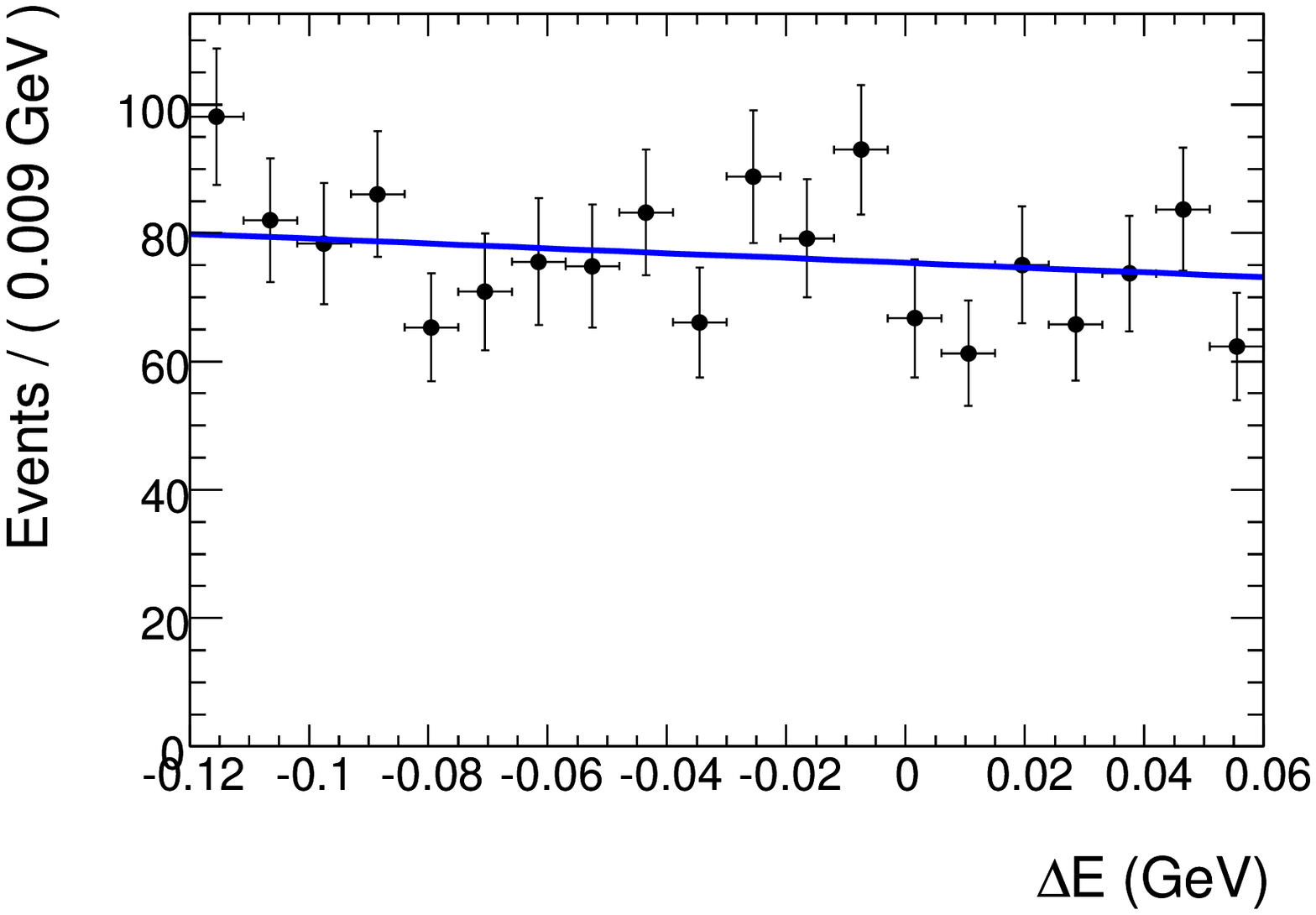}
\end{tabular}
\caption{Distributions of kinematic variables (left) \mes\ and (right) \DeltaE for the \KKKszz subsample: 
(top) signal, (bottom)  continuum background. The points are data events weighted with the \sPlot\ technique~\cite{Pivk:2004ty}, 
and the curves are the PDF shapes used in the ML fit (Sec. \ref{sec:Dalitz}).}
\label{fg::kkks00_event_selection}
\end{figure}

The kinematic variables \mes and \DeltaE are formed for each candidate
as in Sec.~\ref{sec:kkkspm_selection}. Distributions of these
variables are shown in Fig.~\ref{fg::kkks00_event_selection}.
Note that the mean of the signal \DeltaE distribution is shifted from zero due to energy leakage in the EMC.

\subsection{{\boldmath $\Bz \to \Kp\Km\KL$}}
\label{sec:kkklselection}

We identify a \KL candidate 
either as a cluster of energy deposited in the EMC
or as a cluster of hits in two or more layers of the IFR
that cannot be associated with any charged track in the event.
The \KL energy is not measured. Therefore, we determine the \KL
laboratory momentum from its flight direction as measured from the EMC or IFR
cluster, and the constraint that the invariant $\KpKm\KL$ mass equal the
\Bz mass~\cite{Eidelman:2004wy}. In those cases where the \KL is detected in both the IFR and EMC
we use the angular information from the EMC, because it has higher precision.
In order to reduce background from $\pi^0$ decays, we reject
an EMC \KL candidate cluster if it forms an invariant mass between 100
and 150~\mevcc with any other neutral cluster in the event under the $\gamma\gamma$
hypothesis, or if it has energy greater than 1~GeV and contains two shower
maxima consistent with two photons from a $\pi^0$ decay.
The remaining background of \KL candidates due
to photons and overlapping showers is further reduced
with the use of a selector based on the {\it Boosted Decision Trees}
algorithm~\cite{Roe:2004na,Yang:2005nz}.
This selector is constructed from cluster shape variables,
trained with Monte Carlo events, and 
tested on Initial State Radiation $e^+e^-\to\phi(\to\KS\KL )\gamma$ events
and reconstructed $\Bz\to J/\psi \KL$ candidates, which give
a very pure sample of \KL candidates.

We use the kinematic variable $\Delta E$ to characterize the signal 
and background. This variable, computed after the mass constraint on the 
\Bz candidate, has a resolution of about 3.0 \mev for EMC events, 
and about 4.0 \mev for IFR events.
For signal events, \DeltaE\ is expected to peak at zero, with a 
broad tail for positive values of \DeltaE.  We require
$\Delta E < 30$~\mev, in order to be able to 
determine the shape of background under the signal peak. 
The mean value and the resolution of this variable has been taken from 
reconstructed $\Bz\to J/\psi \KL$ events.

In addition to the shape variables described in
Sec.~\ref{sec:selection}, we consider other variables, related to the
missing energy in the event, to characterize events to the \KL final
state. The first, $p^T_{\rm miss}$, is the difference
between the $\Upsilon(4S)$ energy and the total measured energy of the
event, not including the \KL candidate, projected onto the plane transverse
to the beam axis. We also use the angle between
$p^T_{\rm miss}$ and the reconstructed \KL direction, and the
difference between the total visible energy of the event and the
two reconstructed kaon energies. The latter corresponds to the unmeasured \KL\
energy, and has a somewhat different distribution in signal than in
continuum background events.  These three variables are used as
inputs to a Fisher discriminant which has
been trained on Monte Carlo samples and validated on data control samples.

We optimized the selections on all of these variables for maximum
sensitivity in the measurement of \CP parameters. The selection is
optimized independently in the region
$\mKK <1.1~\gevcc$ and in the rest of the Dalitz plot, because
the higher $\Kp\Km$ mass region gets more background from low momentum
neutral candidates, for which the separation between photons and \KL is
worse.  The final average efficiency of the selection is about 25\% in
the low-mass region and 10\% for the rest of events.

\begin{figure}[ptb]
\center
\includegraphics[height=5.5cm]{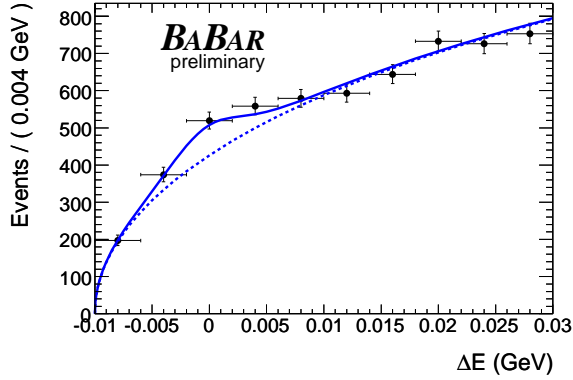}
\caption{Distribution of the kinematic variable \DeltaE for the \KKKl subsample.
The solid line represents the total likelihood, while the dashed line represents the sum of 
continuum and \BB background.
A requirement on the ratio of signal likelihood to signal-plus-background likelihood is 
applied to enhance the signal, with an efficiency of about 30\% for signal.}
\label{fg::kkkl_event_selection}
\end{figure}

The \DeltaE distribution, with the result of the fit superimposed, is 
shown in Fig.~\ref{fg::kkkl_event_selection}, after making a requirement on the ratio 
of signal likelihood to signal-plus-background likelihood to enhance the signal.

\begin{figure}[ptb]
\begin{center}
\begin{tabular}{cc}
\includegraphics[height=5.0cm]{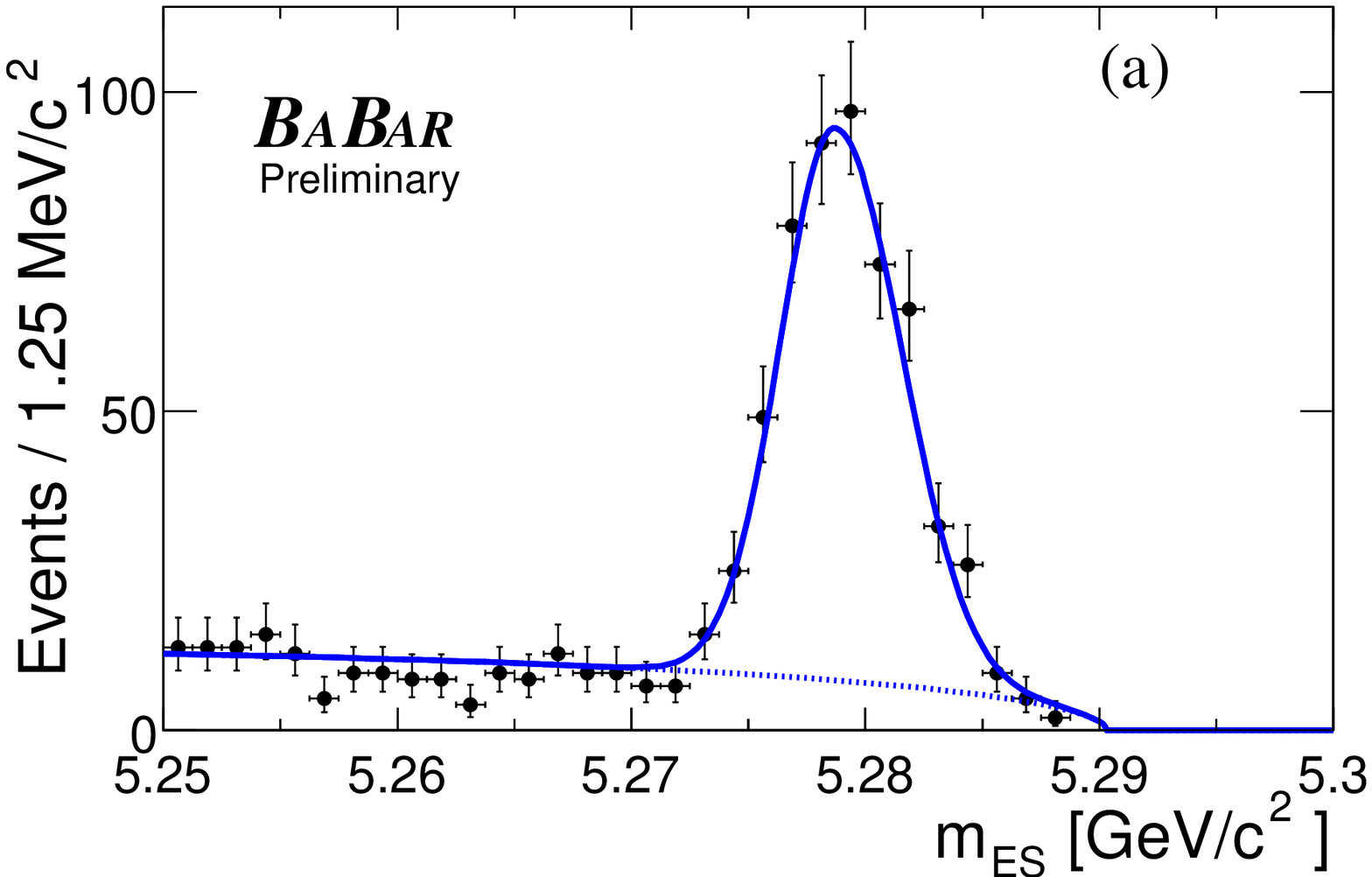} &
\includegraphics[height=5.0cm]{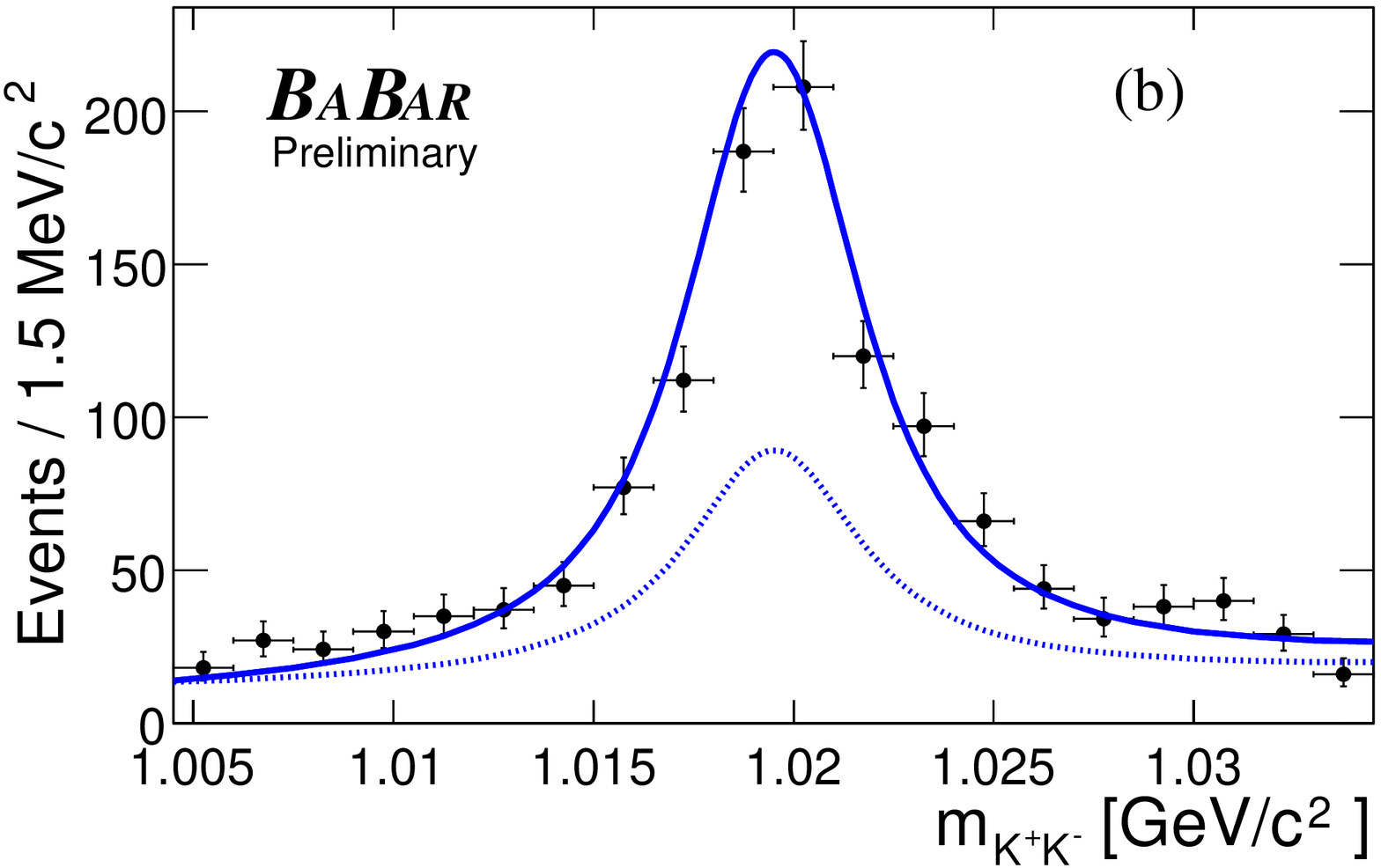} 
\end{tabular} 
\caption{Distributions of the event variables (a) \mes and (b) \mKK
in the $\phi K^+$ final state after reconstruction and a requirement on
the likelihood calculated without the plotted variable. The efficiency
for the selection and likelihood requirements is 78\% for (a) and 95\%
for (b). The solid line represents the fit 
result for the total event yield and the dotted line for the background.
\label{fig:yieldkc}}
\end{center}
\end{figure}

\subsection{{\boldmath $\Bp \to \phi\Kp$}}

We use the $\Bp \to \phi\Kp$ decay to measure the charge asymmetry, 
the $P$-wave fraction, and the relative
phase between the $S$ and $P$ waves in the $\phi(1020)$ region.
The selection of the $\phi$ meson candidate is done by applying an 
invariant $K^+K^-$ mass cut defined as $1.0045 < \mKK < 1.0345$~\gevcc.
For the bachelor $K^+$ candidate from the $B^+$
decay the track requirements are the same as for the $\phi$ daughters
but we apply a more restrictive kaon identification criterion.

The distributions of \mes and \mKK variables after a cut on the ratio
of signal and background probabilities is shown in Figure~\ref{fig:yieldkc}.

%
%

\section{ANALYSIS OF THE DALITZ PLOT}
\label{sec:Dalitz}

We analyze selected events in all neutral $B$ samples using a
maximum likelihood fit with the likelihood function ${\mathcal L}$ 
for each subsample, defined as
\begin{equation}
{\mathcal L} = \exp{\left(-\sum_{i}n_{i}\right)}
\prod_{j}\left[\sum_{i}n_{i}{\mathcal P}_{i,j}\right]
\label{eq::ml}
\end{equation}
where $j$ runs over all events in the sample, and $n_i$ is the yield for event category $i$.
The probability density function (PDF) ${\mathcal P}$ is formed from multiple observables as
\begin{equation}
\label{eq::mlprod}
{\mathcal P} \equiv {\mathcal P}(\mes) \cdot {\mathcal P}(\DeltaE) \cdot {\mathcal P}_{DP}(\mKK, \cos\theta_H, \deltat, q_{tag})\otimes {\cal R}(\deltat, \sigma_{\deltat}).
\end{equation}
Here \cosH is the cosine of the helicity angle between the $\Kp$ and
the $\Kz$ in the $\Kp\Km$ center-of-mass frame, $q_{tag}$ is the
flavor of the initial state, and $\deltat=t_{rec}-t_{tag}$ is the
difference of the proper decay times of the two
$B$-mesons. $\sigma_{\deltat}$ is the error on \deltat. The PDF for
the time-dependent Dalitz plot, ${\mathcal P}_{DP}$, is described in
detail later in the text. $\cal R$ is a standard
\deltat\ resolution function with parameters evaluated in exclusive
$\Bz$ decays into final states with a charm meson as in our
\CP-asymmetry measurements in $J/\psi\KS$ decays~\cite{Aubert:2004zt}.
For the $\Kp\Km\KL$ submode, the \mes variable does not enter the
likelihood function defined in Eq.~(\ref{eq::mlprod}).

In all \KKKz submodes, the signal components of the PDFs for ${\mathcal P}(\mes)$ and ${\mathcal P}(\DeltaE)$ are parameterized using modified Gaussian distributions:
$
{\mathcal P} = \exp [ - (x - x_0)^2/(2 \sigma_{\pm}^2 + \alpha_{\pm} (x- x_0)^2) ],
$
where $x$ is the dependent variable. We determine the parameters $x_0$, $\sigma_+$, $\sigma_-$, $\alpha_+$, and $\alpha_-$ using simulated events, and fix them in fits to data. For
$x<x_0$ ($x>x_0$), the parameters $\sigma_-, \alpha_-$
($\sigma_+, \alpha_+$) are used. In the $\Bp \to \phi \Kp$ mode, the signal \mes distribution is parameterized as above, but with $\alpha_{\pm}$ set to zero. The signal \DeltaE distribution is parameterized as the sum of two Gaussian distributions.

In all applicable submodes, we use the Argus function to model the
continuum background component of ${\mathcal
P}(\mes)$~\cite{Albrecht:1990cs}. For ${\mathcal P}(\DeltaE)$, linear
shapes are used for the continuum background, except for the \KKKl
submode. In that case, we use a reflection of the Argus function.

In the \KKKszz and \KKKl submodes, \BB background components are
parameterized with the same functional forms as the continuum
backgrounds. Due to non-negligible correlation between \mes and
\DeltaE for \BB background in the \KKKszz submode, we construct that
PDF component as a two-dimensional histogram PDF in those variables.

\subsection{Background Decays in the Time-Dependent Dalitz Plot}

The Dalitz plot for the continuum background is parameterized using a
two-dimensional histogram PDF in the variables \mKK and \cosH.  The histogram is filled with
candidates from the region $5.2 < \mes < 5.26~\gevcc$ for the \KKKs
submodes. For the \KKKl submode, candidates from the region $20 <
\DeltaE < 40~\mev$ are used. The \deltat\ distribution is described
with a separate PDF which, similarly to our previous
measurement~\cite{Aubert:2005ja}, uses a double-Gaussian resolution
function and allows a fraction of decays to have a non-zero
lifetime. For the $\Bp \to \phi \Kp$ mode, the continuum background
distribution in
\mKK is modeled with the sum of a relativistic Breit-Wigner function
(see Section \ref{sec:tddp}) and a second-order polynomial.

We estimate the amount of $\BB$ background from Monte Carlo events and again describe the Dalitz plot
using a two-dimensional histogram PDF. The \BB\ background in \KKKs modes is
almost purely combinatorial and is a few percent of the total background. 
The \KKKl sample, in addition to combinatorial \BB\ background,
contains decays into 4-body final states where a pion is missed in the reconstruction (Sec.~\ref{sec:kkklselection}).
Such decays (e.g. $\Kp\Km K^{*}$) are simulated with the assumption that the phase space contains the same
distribution in the Dalitz plot as in $\Bp\to\Kp\Kp\Km$ and $\Bz\to\Kp\Km\KS$ decays: $\phi K^*$,
a wide scalar at 1.5~\gevcc and a non-flat non-resonant distribution.
The \deltat\ distribution is described as a separate PDF that has a non-zero lifetime. 
The time-dependent \CP\ asymmetry of this PDF, set to zero in the reference fit, is varied as a systematic uncertainty.

The decays $\Bz \to \Dp\Km~(\Dp \to \Kp\Kz)$ and $\Bz \to \Ds\Km~(\Ds
\to \Kp\Kz)$ are kinematically indistinguishable from signal decays. We
include non-interfering amplitudes for these modes in our Dalitz plot
model, parameterizing the $D_{(s)}$ mesons on the Dalitz plot as Gaussian
distributions with widths taken from studies of simulated events.

\subsection{Signal Decays in the Time-Dependent Dalitz Plot}
\label{sec:tddp}
When the flavor of the initial state $q_{tag}$, and the difference of
the proper decay times \deltat, are measured,
the time- and flavor-dependent decay rate over the Dalitz plot can be
written as
\begin{eqnarray}
d\Gamma =\frac{1}{(2\pi)^3}\frac{1}{32 M_{\Bz}^3}  \frac{e^{-|\deltat|/\tau_{\Bz}}}{4\tau_{\Bz}} &\times& 
        \Big[~ \left | {\cal A} \right |^2 + \left | \bar{ {\cal A} } \right |^2 \label{eq::DP-q-dt-Rate} 
        \pm q_{tag}  ~2 Im \left (\bar{\cal A} {\cal A}^*  \right ) \sin\deltamd\deltat \\
        && -~ q_{tag} \left (\left | {\cal A} \right |^2 - \left | \bar{ {\cal A} } \right |^2 \right ) \cos\deltamd\deltat 
        ~\Big ], \nonumber
\label{eq::dalitz_plot_rate}
\end{eqnarray}
where plus (minus) sign is for decays to \KKKs (\KKKl) and $q_{tag} = +1~(-1)$ 
when the other \B meson is identified as a \Bz (\Bzb) using a neural network 
technique~\cite{Aubert:2004zt}.
Approximately 75\% of the signal events have tagging information
and contribute to the measurement of CP violation parameters. After
accounting for the mistag rate, the effective tagging efficiency 
is $(30.4\pm 0.3)\%$.
Events without tagging information are still included in the fit as they contribute
to the determination of the Dalitz plot parameters.
Decay amplitudes $\cal A$ and $\bar{\cal A}$ are defined in
(\ref{eq:A}) and (\ref{eq:Abar}) below. 
$M_{\Bz}$, $\tau_{\Bz}$, and \deltamd are
the mass, lifetime, and mixing frequency of the \Bz meson, respectively~\cite{Eidelman:2004wy}.

Four-momentum conservation in a three-body decay gives the relation
$ M^2_{\Bz} + m^2_{1} + m^2_{2} +
m^2_{3} ~=~ m^2_{12} ~+~ m^2_{13} ~+~ m^2_{23}$, where
$m^2_{ij}=(p_i+p_j)^2$ is the square of the invariant mass of a
daughter pair.  This constraint leaves a choice of two independent
Dalitz plot variables to describe the decay dynamics of a spin-zero
particle. In this analysis we choose the $\Kp\Km$ invariant mass \mKK
and the cosine of the helicity angle \cosH.  The PDF for the Dalitz
plot rate becomes
\begin{eqnarray}
        {\cal P}_{DP} \propto d\Gamma(\mKK,\cos\theta_H,\deltat, q_{tag})  \cdot \varepsilon(\mKK,\cos\theta_H)  \cdot |J| 
                        \otimes {\cal R}(\deltat, \sigma_{\deltat}),
\end{eqnarray}
where the Jacobian $|J(\mKK)| = (2 \mKK)(2 q p)$ is given in terms of the charged kaon momentum $q$ and neutral kaon momentum $p$, in the $\Kp\Km$ frame.  
The efficiency $\varepsilon$ is calculated from high-statistics samples of simulated events and depends on the position on the Dalitz plot.

The amplitude $\cal A$ ($\bar{\cal A}$) for the decay $\Bz\to\Kp\Km\Kz$ ($\Bzb\to\Km\Kp\Kzb$) is, in our isobar model, 
written as a sum of decays through intermediate resonances:
\begin{eqnarray}
        {\cal A} &=& \sum\limits_r c_r (1+b_r) e^{i (\phi_r + \delta_r + \beta)} \cdot f_r, \hspace{1cm}\mathrm{and} \label{eq:A} \\
        \bar{\cal A}&=& \sum\limits_r c_r (1-b_r) e^{i (\phi_r - \delta_r - \beta)} \cdot \bar{f}_r. \label{eq:Abar}
\end{eqnarray}
The parameters $c_r$ and $\phi_r$ are the magnitude and phase of the
amplitude of component $r$, and we allow for different isobar
coefficients for $\Bz$ and $\Bzb$ decays through the asymmetry
parameters $b_r$ and $\delta_r$.  The parameter $\beta$ is the CKM
angle $\beta$, coming from \Bz-\Bzb mixing.  The
function $f_r = F_r \times T_r \times Z_r$ describes the dynamic
properties of a resonance $r$, where $F_r$ is the form-factor for the
resonance decay vertex, $T_r$ is the resonant mass-lineshape, and
$Z_r$ describes the angular distribution in the
decay~\cite{blatt,Zemach:1963bc}.

Our model includes the $\phi(1020)$, where we use the  Blatt-Weisskopf centrifugal barrier factor
$F_r=1/\sqrt{1+(Rq)^2}$~\cite{blatt},  where $q$ is the daughter momentum in the resonance frame,
and $R$ is the effective meson radius, taken to be $R=1.5~\gev~(0.3~\fm)$. 
For the scalar decays included in our model ($f_0(980)$, $X_0(1550)$, and $\chi_{c0}$), we use a constant form-factor.
Note that we have omitted a similar centrifugal factor for the $\Bz$ decay vertex into the $\phi\Kz$ intermediate state
since its effect is negligible due to the small width of the $\phi(1020)$ resonance.

The angular distribution is constant for scalar decays, whereas for vector decays $Z=-4 \vec{q} \cdot \vec{p}$, 
where $\vec{q}$ is the momentum of the resonant daughter, and $\vec{p}$ is the momentum of the third particle in 
the resonance frame. 
We describe the line-shape for the $\phi(1020)$, $X_0(1550)$, and $\chi_{c0}$  using the relativistic Breit-Wigner function 
\begin{equation}
        T(m) = \frac{1}{m^2_r - \mKK^2 - i m_r \Gamma(m)},
\end{equation}
where $m_r$ is the resonance pole mass. The mass-dependent width is given as
$
\Gamma(\mKK) =\Gamma_r  \left ( q/q_r\right )^{2L+1} \left ( m_r / \mKK  \right ) \left ( F_r(q)/F_r(q_r) \right )^2,
$
where $L$ is the resonance spin and $q=q_r$ when $\mKK=m_r$. For the $\phi(1020)$ and $\chi_{c0}$ parameters, we
use average measurements~\cite{Eidelman:2004wy}. The $X_0(1550)$ is less well-established.
Previous Dalitz plot analyses of $\Bp\to\Kp\Kp\Km$~\cite{Garmash:2004wa,Aubert:2006nu} and
$\Bz\to\Kp\Km\Kz$ decays~\cite{Aubert:2005kd}  report observations of a scalar resonance at around 1.5~\gevcc.
The scalar nature has been confirmed by partial-wave analyses~\cite{Aubert:2005ja,Aubert:2006nu}.
However, previous measurements report inconsistent resonant widths: $0.145\pm 0.029$~\gevcc~\cite{Garmash:2004wa} and
$0.257 \pm 0.033$~\gevcc~\cite{Aubert:2006nu}. Branching fractions also disagree, so the nature of this component is still unclear~\cite{Minkowski:2004xf}.
In our reference fit, we take the resonance parameters from Ref.~\cite{Aubert:2006nu}, which is based on
a larger sample of $\BB$ decays than Ref.~\cite{Garmash:2004wa}, and consider the narrower width given in the latter in the systematic error studies.

The $f_0(980)$ resonance is described with the coupled-channel (Flatt\'e) function
\begin{equation}
        T(\mKK)    =  \frac{1 }{   m^2_r - \mKK^2 - i m_r ( \rho_{K} g_{K} + \rho_{\pi} g_{\pi}  )     }, 
\end{equation}
where  $\rho_K (\mKK) =2 \sqrt{ 1 - 4 m^2_{K}/\mKK^2 }$, $\rho_\pi (\mKK) =2 \sqrt{ 1 - 4 m^2_{\pi}/\mKK^2 }$, and
the coupling strengths for the $KK$ and $\pi\pi$ channels are 
taken as $g_\pi=0.165\pm0.018$~\gevcc, $g_K/g_\pi=4.21 \pm 0.33$, 
and $m_r=0.965 \pm 0.010 $~\gevcc~\cite{Ablikim:2004wn}.

In addition to resonant decays, we include non-resonant amplitudes. 
Existing models consider contributions from contact terms or higher-resonance
tails~\cite{Cheng:2002qu,Fajfer:2004cx,Cheng:2005ug}, but they do not capture features observed in data.
We rely on a phenomenological parameterization~\cite{Garmash:2004wa} and 
describe the non-resonant terms as
\begin{equation}
         {\cal A}_{NR} =  \left ( c_{12} e^{i\phi_{12}} e^{-\alpha m^2_{12}}  
             +   c_{13} e^{i\phi_{13}} e^{-\alpha m^2_{13}}  
             +    c_{23} e^{i\phi_{23}} e^{-\alpha m^2_{23}} \right )  
	     \cdot (1+b_{NR}) \cdot  e^{i(\beta+\delta_{NR})} \label{eq:nr}
\end{equation}
and similarly for $\bar{\cal A}_{NR}$.
The slope of the exponential function is consistent among previous measurements in both neutral and
charged $B$ decays into three kaons~\cite{Garmash:2004wa,Aubert:2006nu,Aubert:2005kd}, and we use $\alpha = 0.14 \pm 0.02~\gev^{-2} \cdot c^4$.

We compute the direct \CP-asymmetry parameters for resonance $r$ from the asymmetries in amplitudes ($b_r$) 
and phases ($\delta_r$) given in Eqs.~(\ref{eq:A}, \ref{eq:Abar}). We define the rate asymmetry as
\begin{equation}
        \Acp(r)=\frac{|\bar{\cal A}_r|^2 - |{\cal A}_r|^2}{|\bar{\cal A}_r|^2 + |{\cal A}_r|^2} =\frac{-2b_r}{1+b_r^2},
\label{eq:Acp}
\end{equation}
and $\betaeff (r) = \beta + \delta_r$ is defined as the total phase asymmetry.
The fraction for resonance $r$ is computed 
\begin{equation}
	{\cal F}_r ~=~ \frac{ \int d\cos\theta_H ~d\mKK \cdot |J| \cdot (|{\cal A}_r |^2+|\bar{\cal A}_r|^2)    }
		     { \int  d\cos\theta_H ~d\mKK \cdot |J| \cdot (|{\cal A} |^2+|\bar{\cal A}|^2)    }.
\end{equation}
The sum of the fractions can differ from unity due to interference between the isobars.

\subsection{Calculation of Angular Moments}

As an alternative description that is less dependent on a
resonance model, we analyze the Dalitz plot in terms of moments of
the cosine of the  helicity angle, \cosH, in $\Bp \to \phi\Kp$ and $\Bz \to \KKKspm$ decays.
We only assume that the total amplitude is the sum of the two lowest partial waves, and ignore direct \CP-violation effects:
\begin{equation}
        {\cal A}~(\bar{\cal A})  ~=~ A_s P_0(\cos\theta_H) ~\pm~ e^{i\phi} A_p P_1(\cos\theta_H),
\end{equation}
where $\phi = \phi_P-\phi_S$ is the relative phase between $S$ and $P$-wave strengths.
The $\pm$ sign corresponds to $\Bz$ or $\Bzb$ decays, respectively.
$P_0$ and $P_1$ are Legendre polynomials that describe the amplitude dependence on \cosH for $S$-wave and $P$-wave decays, respectively.
Integrating Eq.~(\ref{eq::dalitz_plot_rate}) over \deltat, the tag-dependent decay rate can be written in terms of Legendre polynomials as follows:
\begin{eqnarray}
        \frac{d\Gamma}{d\cos\theta_H \cdot d\mKK \cdot |J|}   &=& \sum\limits_{l=0,1,2}~\left < P_l \right > \times P_l(\cos\theta_H) \\
                        &=&
                        \frac{A_s^2 + A_p^2}{\sqrt{2}} \times P_0(\cos\theta_H) ~+~ \sqrt{\frac{2}{5}} A_p^2 \times P_2(\cos\theta_H) \nonumber \\ 
                        & & - \frac{q_{tag} \cdot \left < D \right >}{(\deltamd \tau_{\Bz})^2+1} \times
                                 \frac{ 2 A_s A_p}{\sqrt{2}} \cos\phi \times P_1(\cos\theta_H), 
\label{eq::Legendre_moments_rate}
\end{eqnarray}
where the flavor tagging is necessary to measure the $S$-$P$ wave interference term proportional to $P_1$.
In the decay $\Bz\to\KKKz$, this term is diluted by the imperfect tagging $\left < D \right >$ and the $\Bz-\Bzb$ mixing.
In charged $B$ decays to the $\Kp\Kp\Km$ final state we  set $\deltamd=0$ and  $\left < D \right >=1$ since $q_{tag}$ corresponds
to the charge of the final state.
Using our data sample, we compute the Legendre moments $\left < P_l \right  >$,
\begin{eqnarray}
        \left < P_l \right > &\approx& \sum_i P_l(\cos\theta_H, i) ~{\cal W}(i) / \varepsilon(i),
\label{eq::Legendre_moments_data}
\end{eqnarray}
where $\cal W$ is the weight for event $i$ to belong to the signal sample~\cite{Pivk:2004ty}.
These weights are computed from maximum likelihood fits that do not use the mass or the helicity angle
in the fit. Finally, the fraction of P-wave decays is computed as
$       f_{p} = \sqrt{5/4} \cdot \left < P_2 \right > / \left < P_0 \right > $~\cite{Aubert:2005ja}.

%
%

\section{RESULTS}
\label{sec:Physics}

\subsection{Dalitz Plot and Angular Moments}

In order to determine parameters of the Dalitz plot model,
we perform a fit to 3091 $\Bz\to\KKKspm$  candidates  in the full Dalitz plot.
In this step we assume that all decays have the same \CP-asymmetry parameters.
We vary the event yields, isobar coefficients of the Dalitz plot model, and two
\CP-asymmetry parameters averaged over the Dalitz plot. 
We find a signal yield of $879 \pm 36$ events.
The isobar amplitudes, phases, and fractions are listed in Table~\ref{tab:isobars}.
The sum of resonant fractions in our DP model is different
from 100\% due to interference between resonances.

We compare our fractions with other Dalitz plot analyses using flavor symmetry~\cite{Gronau:2005ax}.
We find consistent fractions for decays through the $\phi(1020)$ resonances with the 
$\Bp\to\Kp\Kp\Km$ decay~\cite{Garmash:2004wa,Aubert:2006nu}.
The fraction of $f_0(980)\Kz$ decays is consistent with our $\Bp\to\Kp\Kp\Km$ analysis,
and all $\Bp\to\Kp\pip\pim$ Dalitz plot analyses~\cite{Garmash:2004wa,Aubert:2006nu,Aubert:2005ce}.
The fraction of non-resonant decays, which is predicted to be half
of the contribution in  $\Bp\to\Kp\Kp\Km$~\cite{Gronau:2005ax}, is harder to compare since 
existing measurements in the charged mode are inconsistent. Our result agrees well with 
\babar's result~\cite{Aubert:2006nu}, and is within two standard deviations of Belle's result~\cite{Garmash:2004wa}.
Determination of the wide scalar resonance at 1.5~\gevcc, labeled as $X_0(1550)$, is even more uncertain.
Using the same resonant parameters we find a much smaller fraction than in \babar's analysis~\cite{Aubert:2006nu},
but our solution is more consistent with Belle's $\Bp\to\Kp\Kp\Km$ analysis~\cite{Garmash:2004wa}.

\begin{figure}[ptb]
\center
\begin{tabular}{ll}
a)\\
\includegraphics[height=5.5cm]{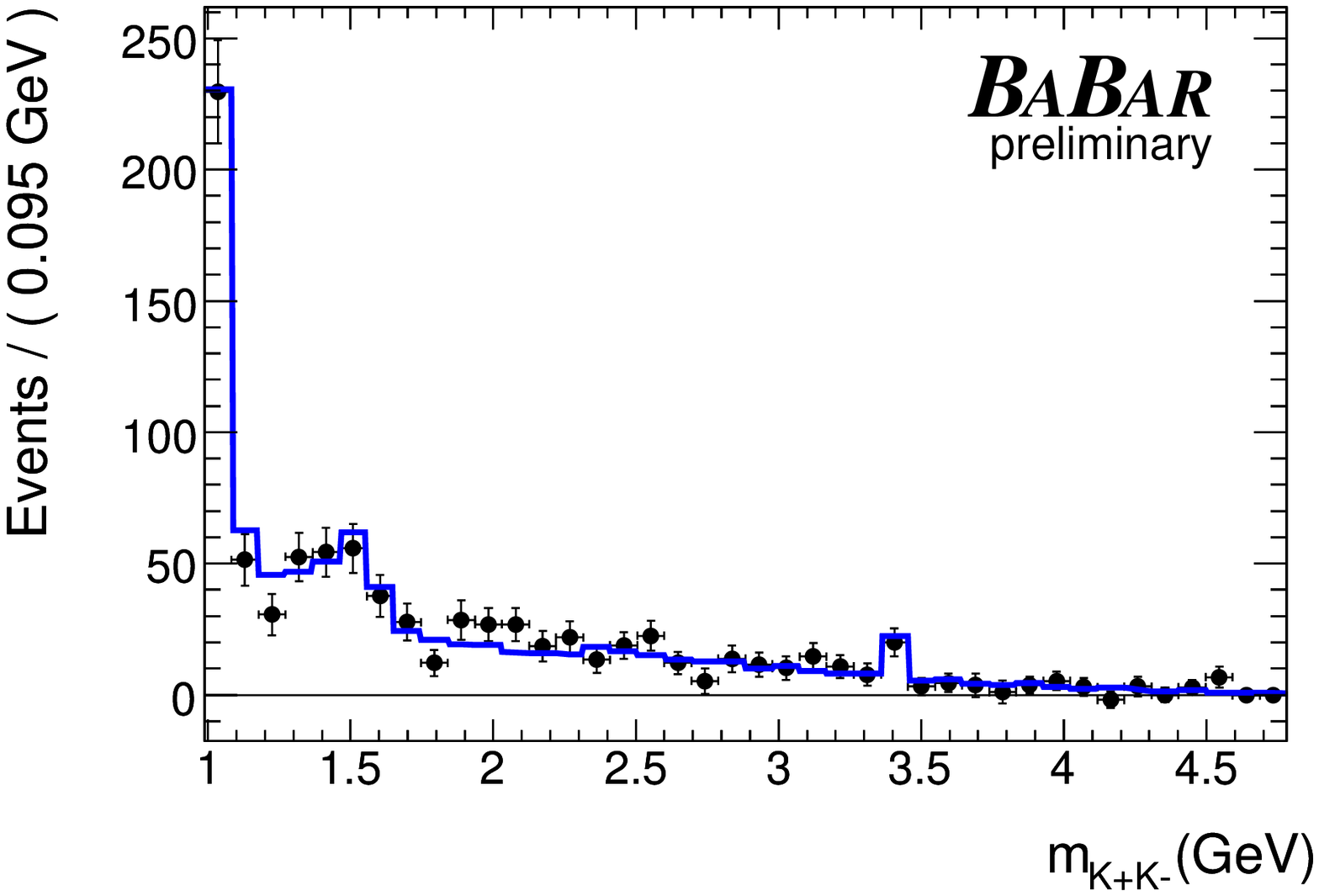} & \includegraphics[height=5.5cm]{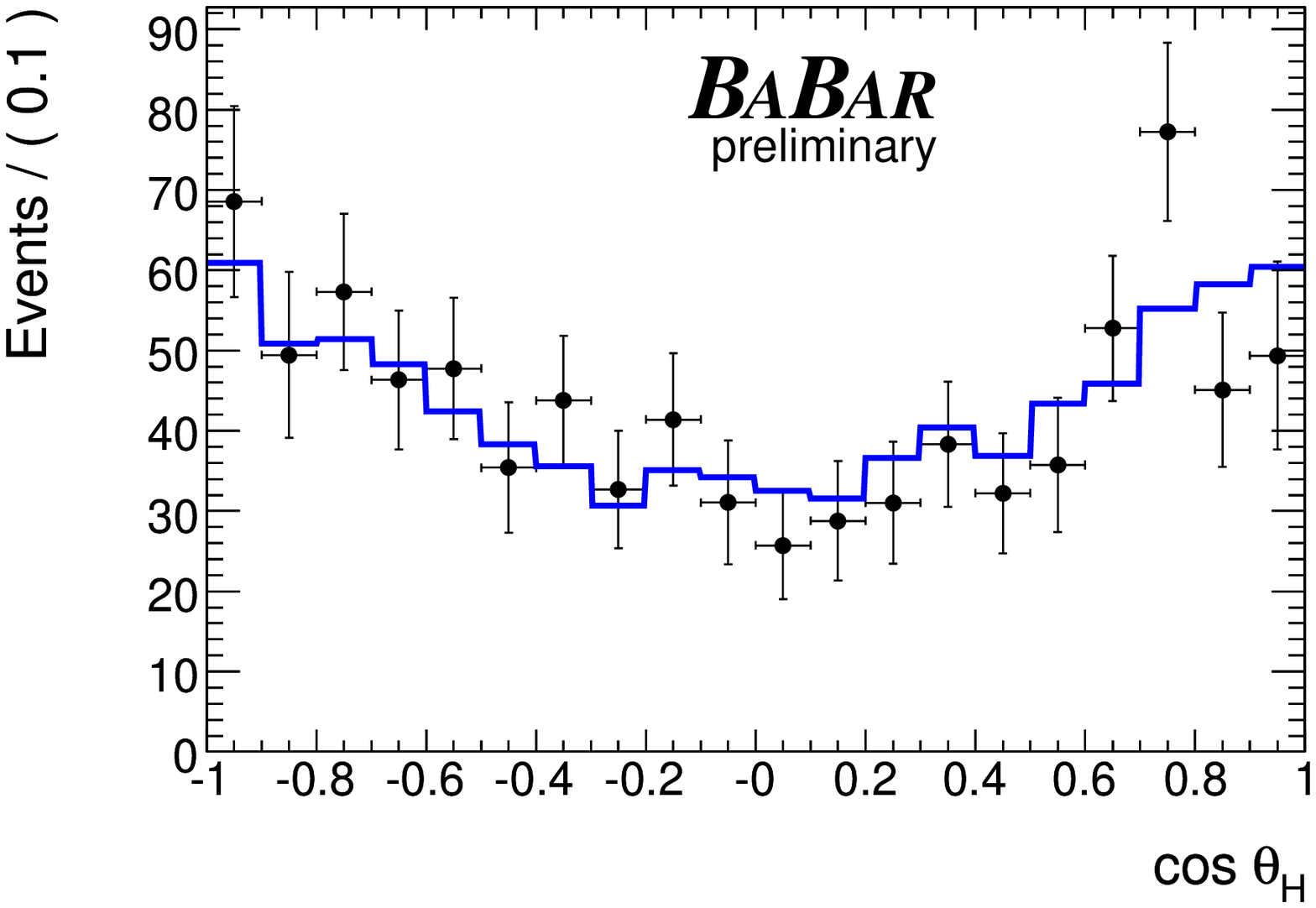} \\
b)\\
\includegraphics[height=5.5cm]{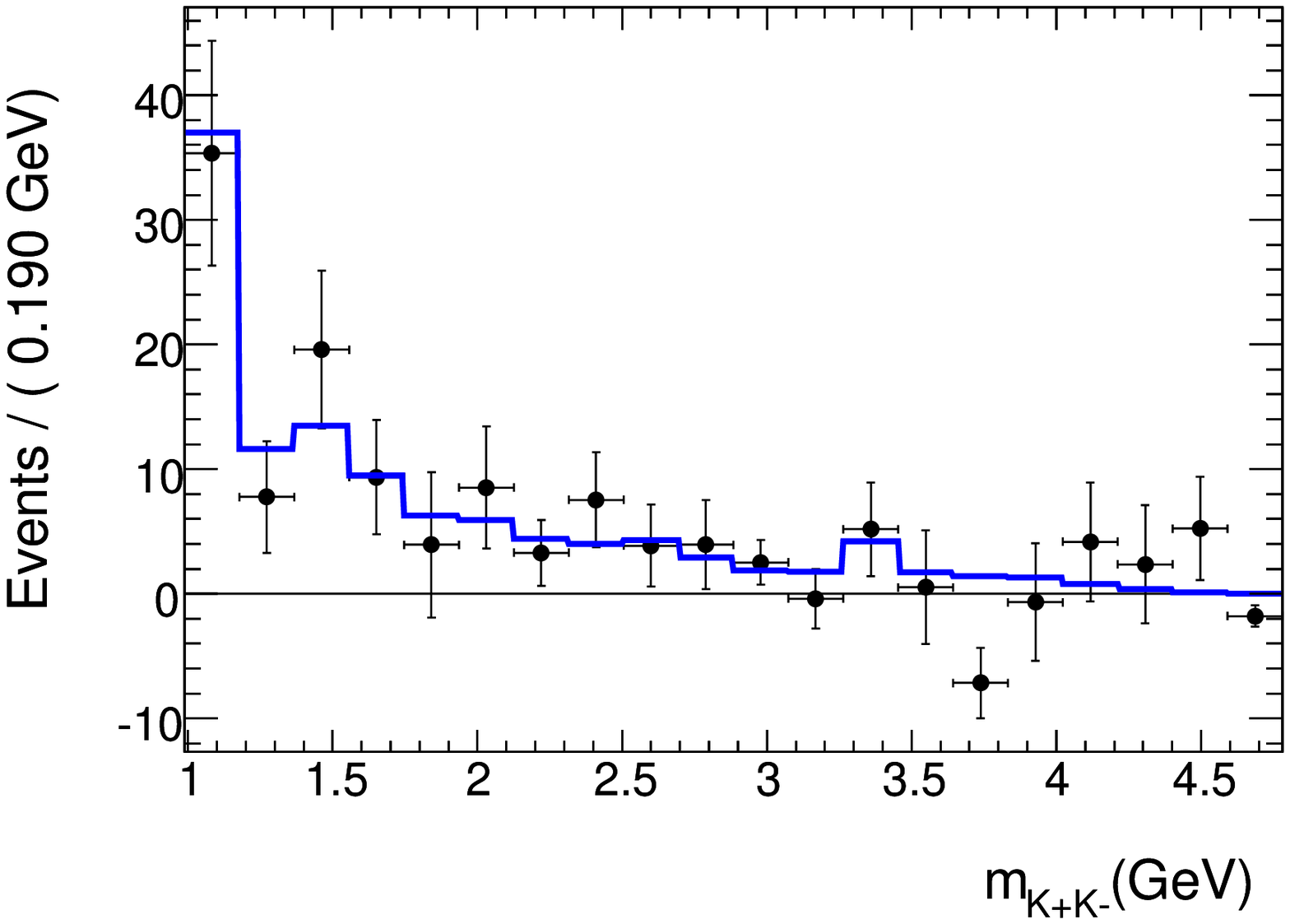} & \includegraphics[height=5.5cm]{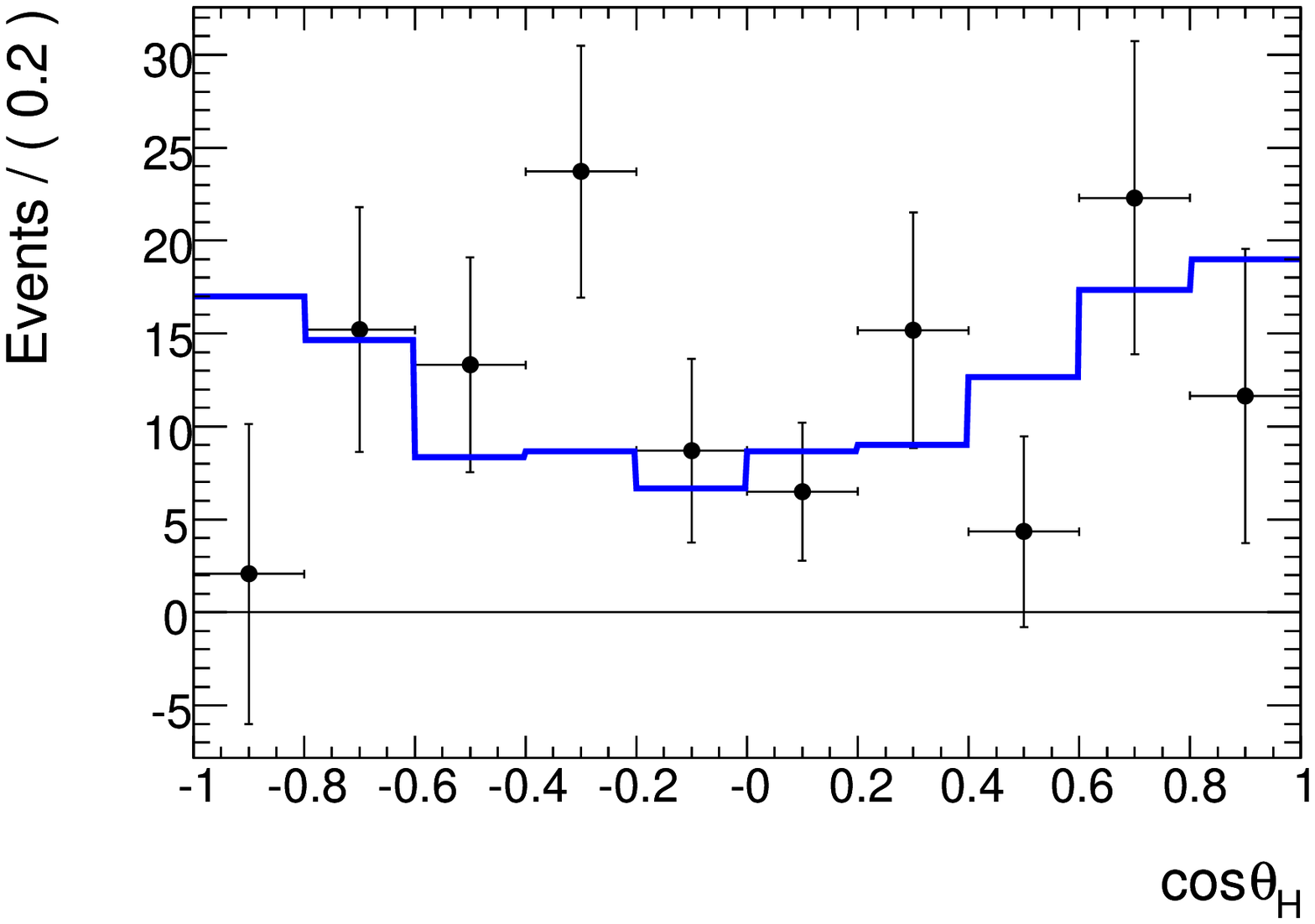} \\
c)\\
\includegraphics[height=5.5cm]{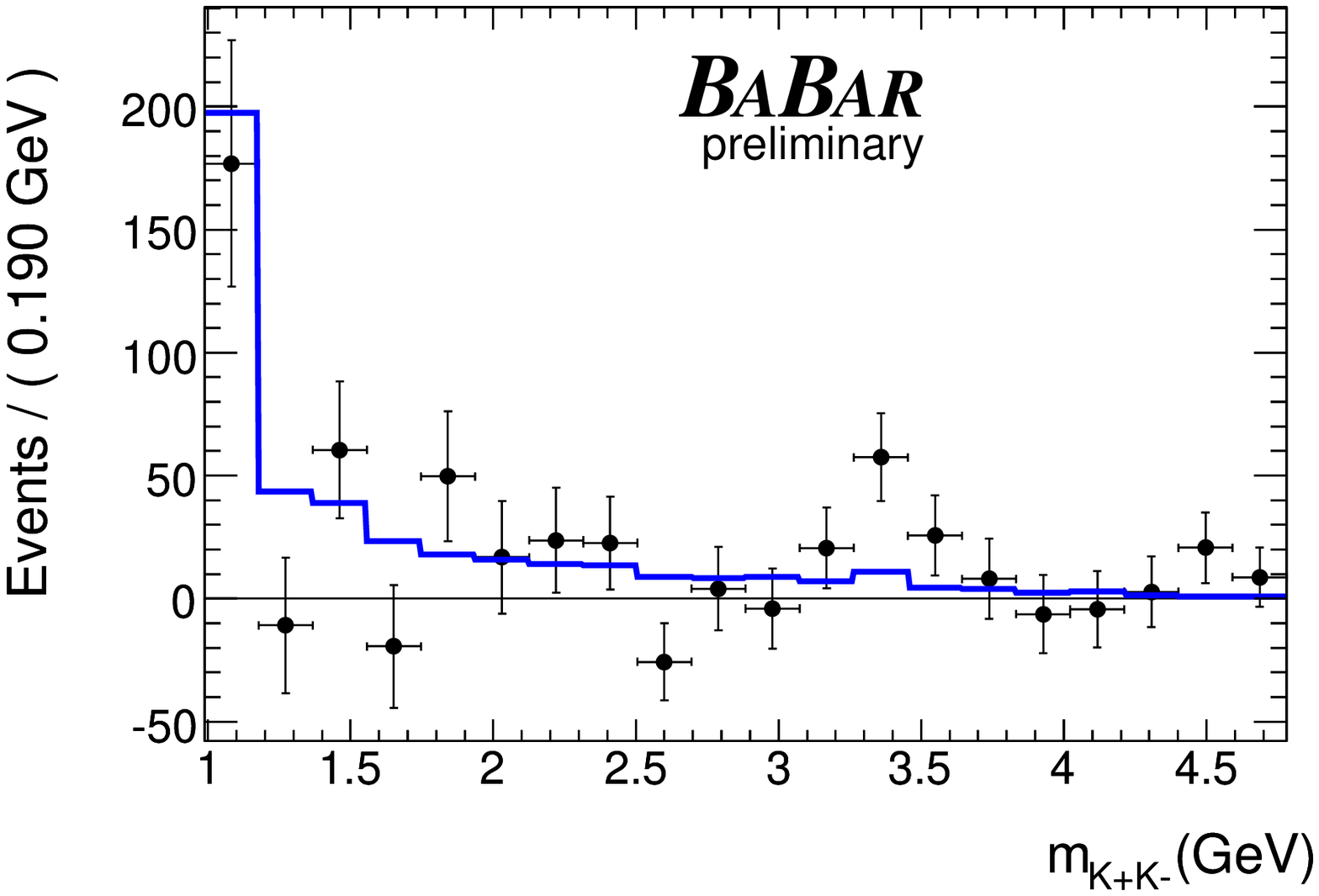} & \includegraphics[height=5.5cm]{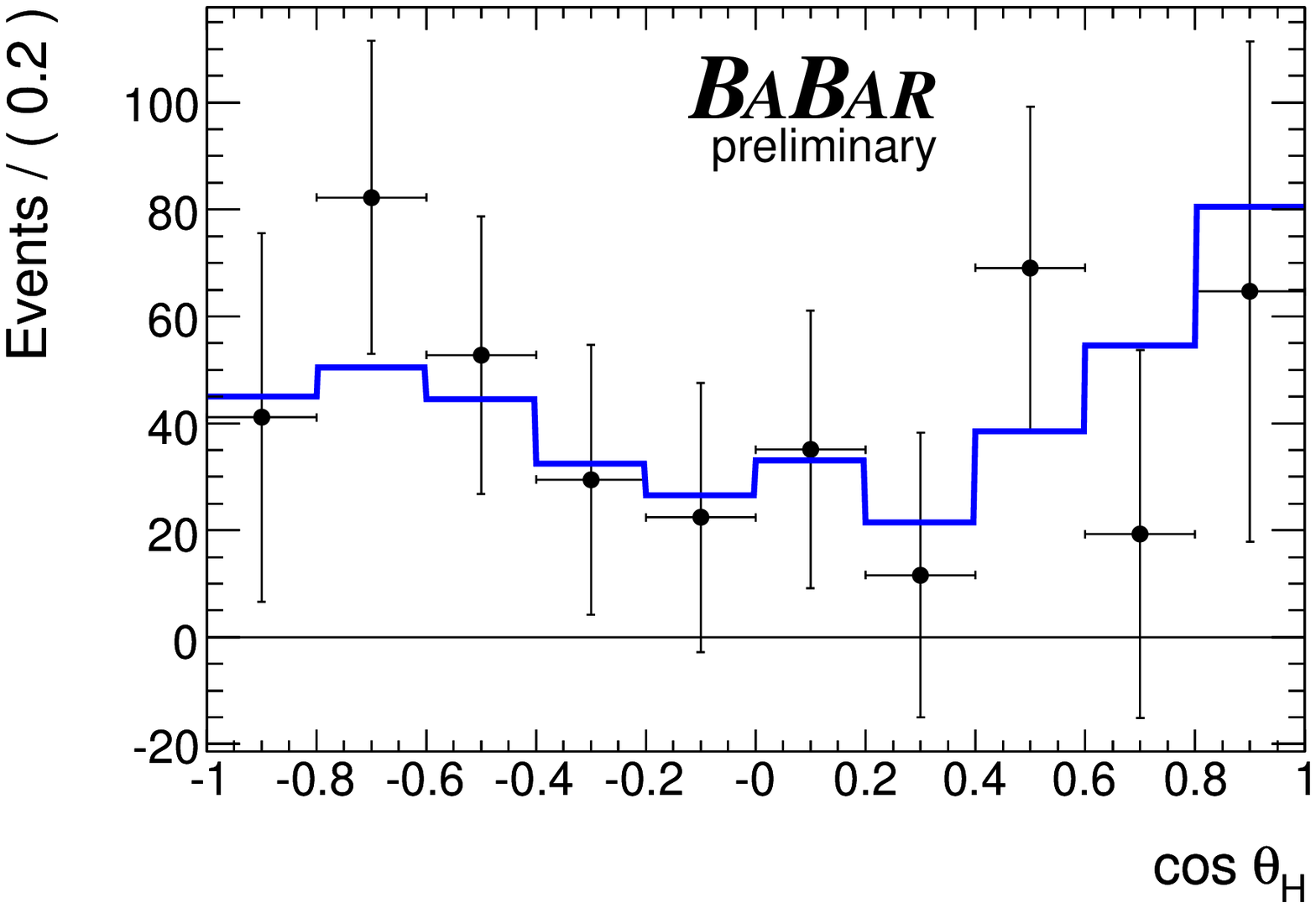} \\
\end{tabular}
\caption{Distributions of the Dalitz plot variables (left) \mKK and (right) \cosH for signal events (points) 
compared with the fit PDF in the following sub-samples: a) \KKKspm, b) \KKKszz, c) \KKKl. All sub-samples
use the same Dalitz plot model but have different efficiencies so the resulting distributions differ.}
\label{fg::dalitz-splot_all-same-dcpv}
\end{figure}

In order to achieve better statistical precision in measurements of \CP-asymmetry parameters that are
described in the following sections, we combine the \KKKspm sample with samples of 1599 
$\Bz\to\KKKszz$ and 22341 
$\Bz\to\KKKl$ candidates. 
Fixing the coefficients of the isobar model to the values extracted from the \KKKspm submode, 
we find signal yields of $138 \pm 17$ and $499 \pm 52$ events in the
\KKKszz and \KKKl submodes, respectively. 
Projection plots for the Dalitz plot variables
are shown for all three submodes in Figure~\ref{fg::dalitz-splot_all-same-dcpv}.

\begin{table}[h]
\center
\begin{tabular}{|lr|rrr|}
\hline \hline
\multicolumn{2}{|l|}{Decay }            &     Amplitude $c_r$   	&       Phase $\phi_r$   &       Fraction ${\cal F}_r$ (\%)\\
\hline  
\multicolumn{2}{|l|}{$\phi(1020)\Kz$}	& $  0.0098\pm 0.0016$       &       $ -0.11 \pm 0.31$   &       $12.9 \pm 1.3$         \\ 
\multicolumn{2}{|l|}{$f_0(980)\Kz$}	& $  0.528 \pm 0.063$         &       $ -0.33 \pm 0.26$       &       $22.3 \pm 8.9$    \\
\multicolumn{2}{|l|}{$X_0(1550)\Kz$}	& $  0.130 \pm 0.025$  		&       $ -0.54 \pm 0.24$   &       $4.1 \pm 1.8$      \\ 
$NR$ &$(\Kp\Km)$      			& 1 (fixed)             	&       0 (fixed)        &                              \\
        &$(\Kp\Kz)$             	& $  0.38 \pm 0.11$      	&       $  2.01  \pm 0.28$     &       $91 \pm 19$     \\
        &$(\Km\Kz)$             	& $  0.38 \pm 0.16$   	&       $ -1.19  \pm 0.37$    &                           \\
\hline
\multicolumn{2}{|l|}{$\chi_{c0}\Kz$}    & $  0.0343\pm 0.0067$  	&       $ 1.29 \pm 0.41$    &       $2.84 \pm 0.77$       \\
\multicolumn{2}{|l|}{$\Dp\Km$}          & $  1.18\pm 0.24$    	&               --       	&       $3.18 \pm 0.89$          \\
\multicolumn{2}{|l|}{$\Ds\Km$}          & $  0.85\pm 0.20$     	&               --          &       $1.72 \pm 0.65$        \\
\hline \hline
\end{tabular}
\caption{Isobar amplitudes, phases, and fractions from the fit to the $\Bz\to\KKKspm$ sample. 
Three rows for non-resonant contribution correspond to coefficients of exponential functions in Eq.~(\ref{eq:nr}), 
while the fraction is given for the combined amplitude. Errors are statistical only. }
\label{tab:isobars}
\end{table}

As an additional crosscheck of our Dalitz plot model,
we compute angular moments and extract strengths of the partial waves in $\Kp\Km$ mass bins using the $\Bp\to\phi\Kp$ 
and $\Bz \to \KKKspm$  samples. In this approach we rely only on the assumption that the two lowest partial waves are present,
but make no other assumption on the decay model. 
We confirmed the non-existence of higher partial 
waves by determining that higher angular moments ($\left <P_{3-5}\right >$) are consistent with zero.

\begin{figure}[ptb]
\begin{center}
\begin{tabular}{cc}
\includegraphics[height=6.5cm]{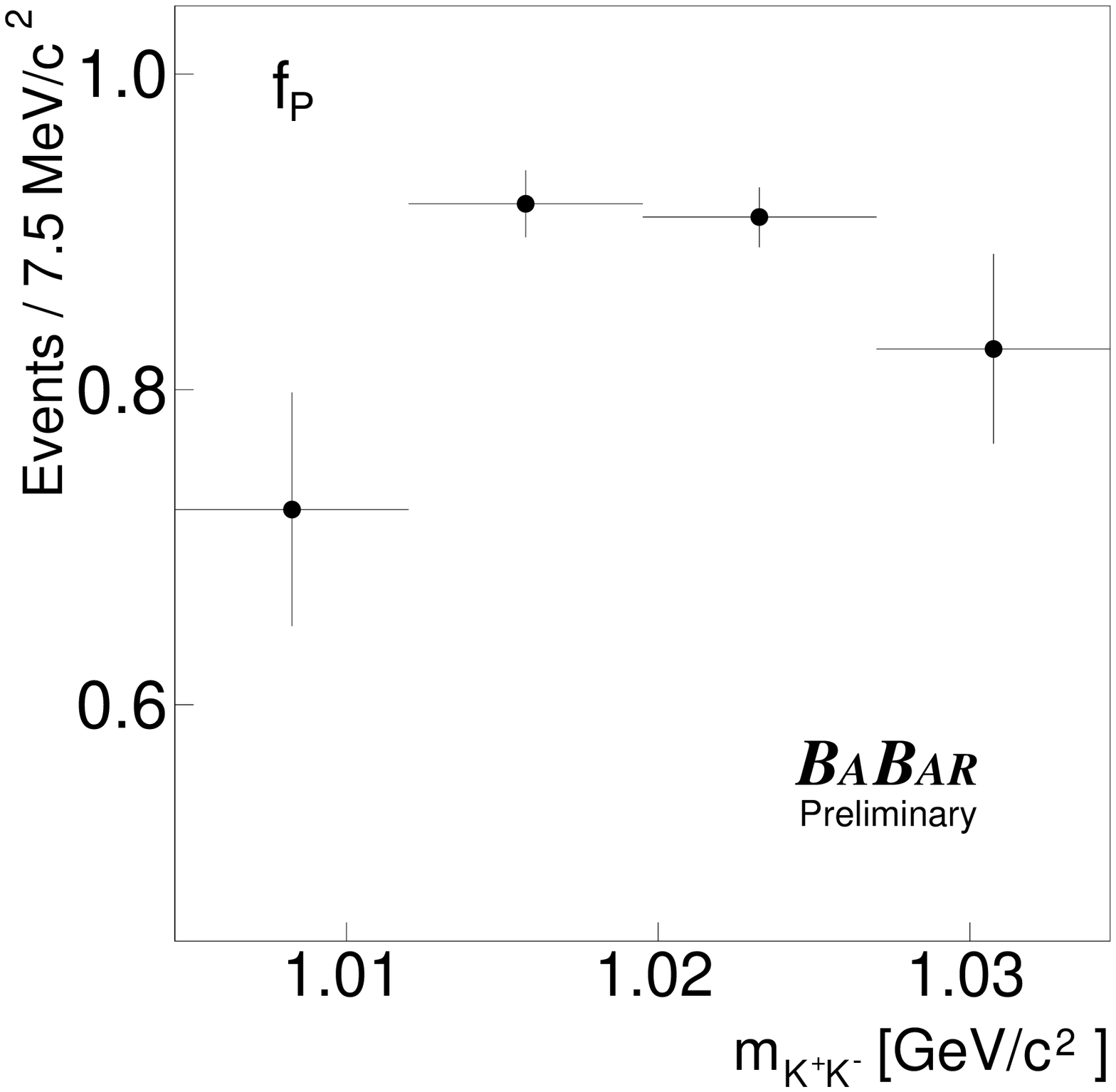} &
\includegraphics[height=6.5cm]{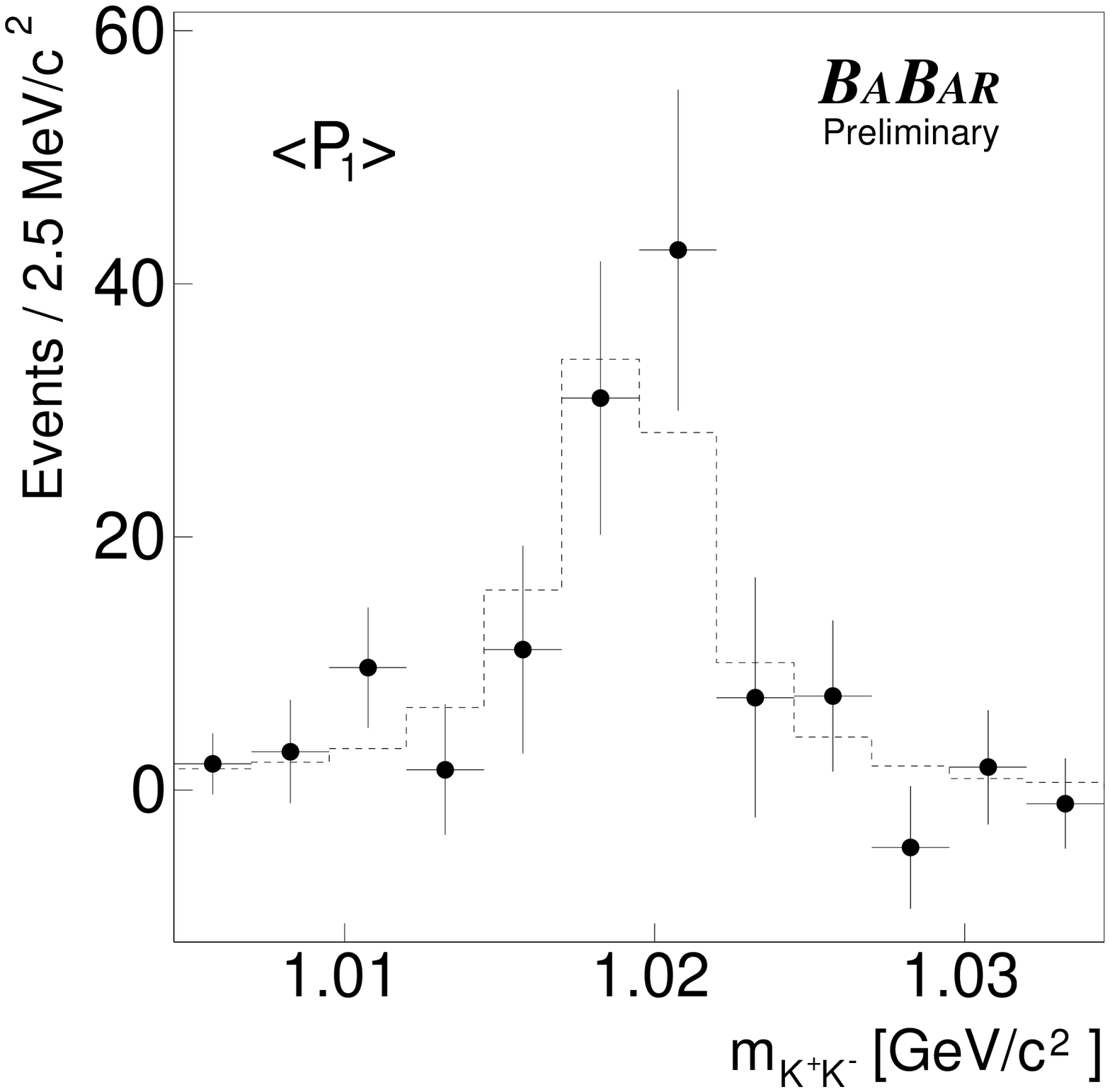} 
\vspace{-.5cm}
\end{tabular} 
\caption{$\phi K^+$: (left) The relative $P$-wave fraction $f_p$ in the interval 
$1.0045 < \mKK < 1.0345$~\gevcc.
(right) The moment $\left < P_1 \right >$ calculated with the $\phi(1020)$ helicity angle
defined with respect to the kaon of the same charge as the $B$ meson. The dashed line corresponds 
to the fit result.
\label{fig:fracfp}}
\end{center}
\end{figure}

From the fit to 4947 $\Bp\to\phi\Kp$ candidates, we find $624\pm 30$ signal
candidates in the mass region $1.0045 < \mKK < 1.0345$~GeV/c$^2$.
The event weights ${\cal W}_i$ are computed from the likelihood
without \mKK and \cosH. From $\left < P_0 \right >$ and $\left
< P_2 \right >$ we obtain the average fraction $f_p = 0.891\pm 0.014$.
The distribution of $f_p$ in four bins of \mKK is shown in
Fig.~\ref{fig:fracfp}.  In order to determine the relative phase
between $S$- and $P$-waves, we make a $\chi^2$ minimization of the
moment $\left < P_1 \right >$ given with
Eq.~(\ref{eq::Legendre_moments_rate}), by varying the phase while
keeping $S$- and $P$-wave strengths fixed. For the $S$-wave we assume
negligible energy dependence in the $\phi(1020)$ region and
parameterize the resonance shape with a relativistic Breit-Wigner
function.  We estimate a phase difference between $P$- and $S$-wave of
$(78 \pm 20)^{\circ}$; the systematic error due to the choice of the
$S$-wave model and the $\chi^2$ scan method is negligible. The moment
$\left < P_1 \right >$ from data, with the best phase solution
superimposed, is displayed in the right-hand plot of Figure~\ref{fig:fracfp}.

\begin{figure}[ptb]
\begin{center}
\begin{tabular}{c}
\includegraphics[height=6.5cm]{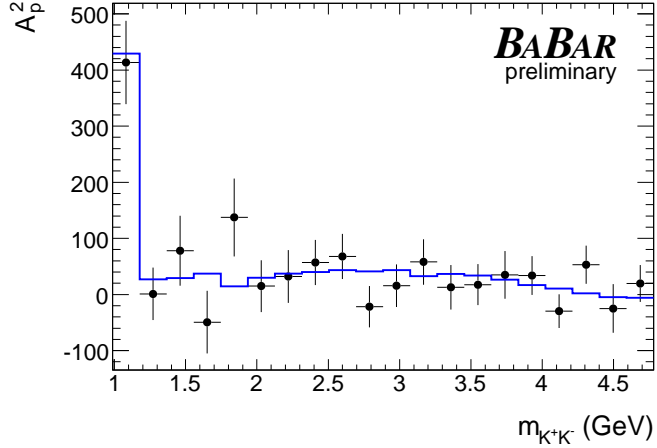} 
\vspace{-.5cm}
\end{tabular} 
\caption{\KKKspm: The absolute strength of $P$-wave decays as a function of $\Kp\Km$ mass.
The points are signal-weighted data and the histogram corresponds to the Dalitz plot model.
\label{fig:kkkspm_P-wave-strength}}
\end{center}
\end{figure}

Similarly, we compute the $P$-wave strength as a function of $\Kp\Km$
invariant mass in $\Bz \to \KKKspm$ decays as shown in
Figure~\ref{fig:kkkspm_P-wave-strength}.  We find the total fraction
of $P$-wave in our sample is $0.29 \pm 0.03~\mathrm{(stat)}$
integrated over the entire Dalitz plot, which is consistent with our
previous measurement~\cite{Aubert:2005ja}. In our model, this $P$-wave
contribution originates from $\phi(1020)\KS$ decays, and non-resonant
events with $\Kp\KS$ and $\Km\KS$ mass dependence that reflects into
an effective P-wave that enhances the central part of the plot in
Figure~\ref{fig:kkkspm_P-wave-strength}.

\subsection{{\boldmath $\CP$} Asymmetry in the Low-{\boldmath $\Kp\Km$} Mass Region}

\begin{figure}[ptb]
\center
\begin{tabular}{ll}
a)\\
\includegraphics[height=5.5cm]{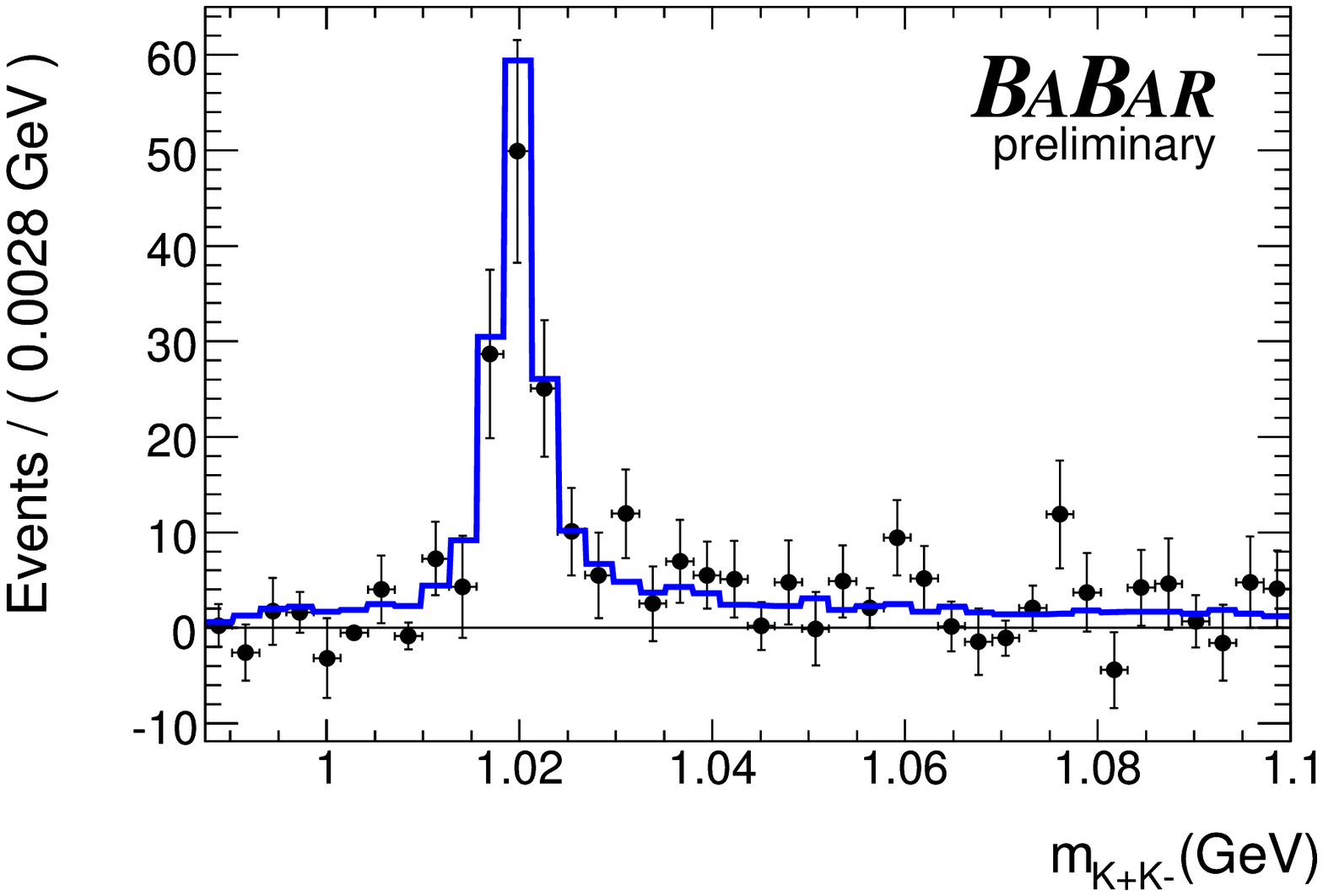} & \includegraphics[height=5.5cm]{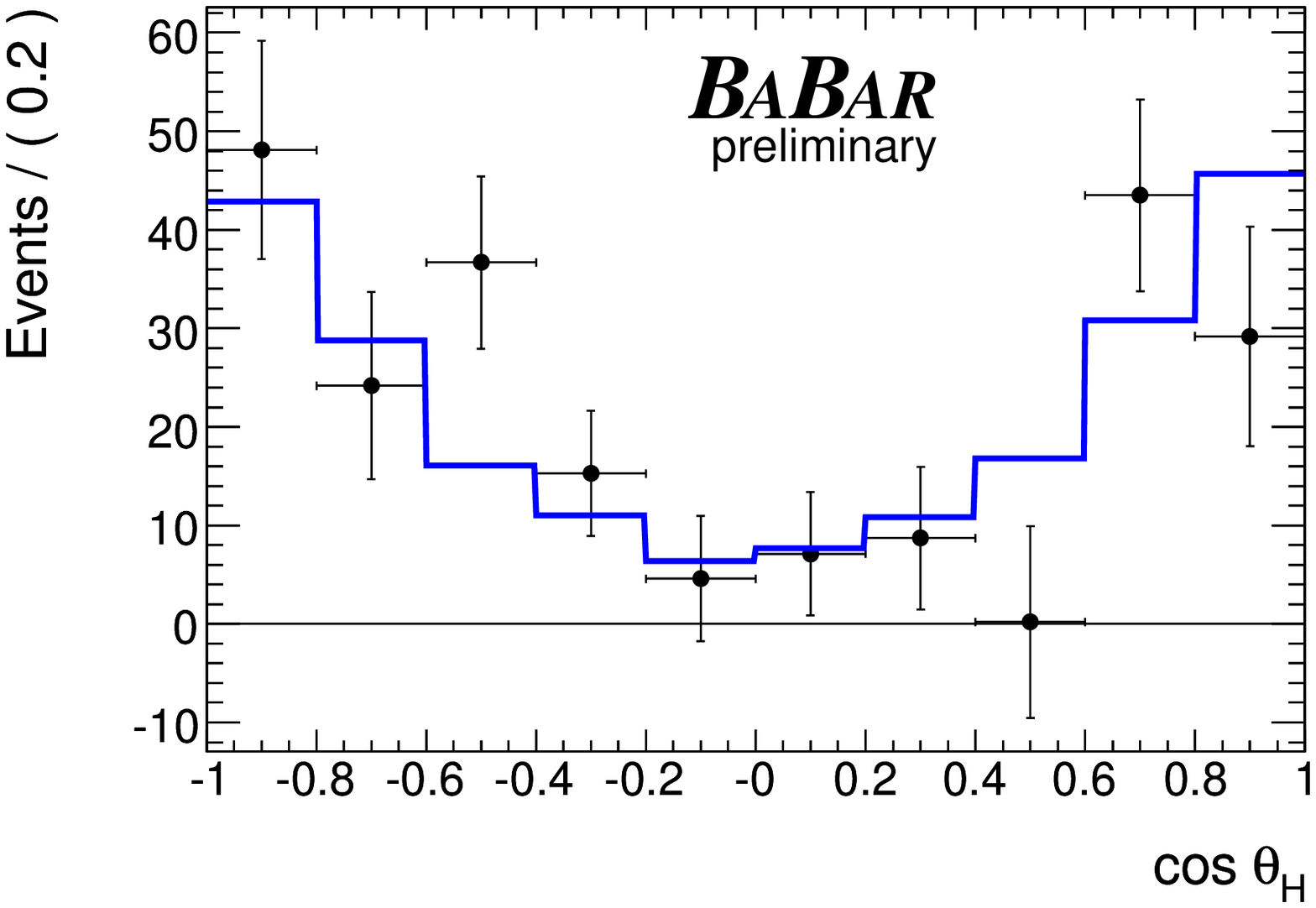} \\
b)\\
\includegraphics[height=5.5cm]{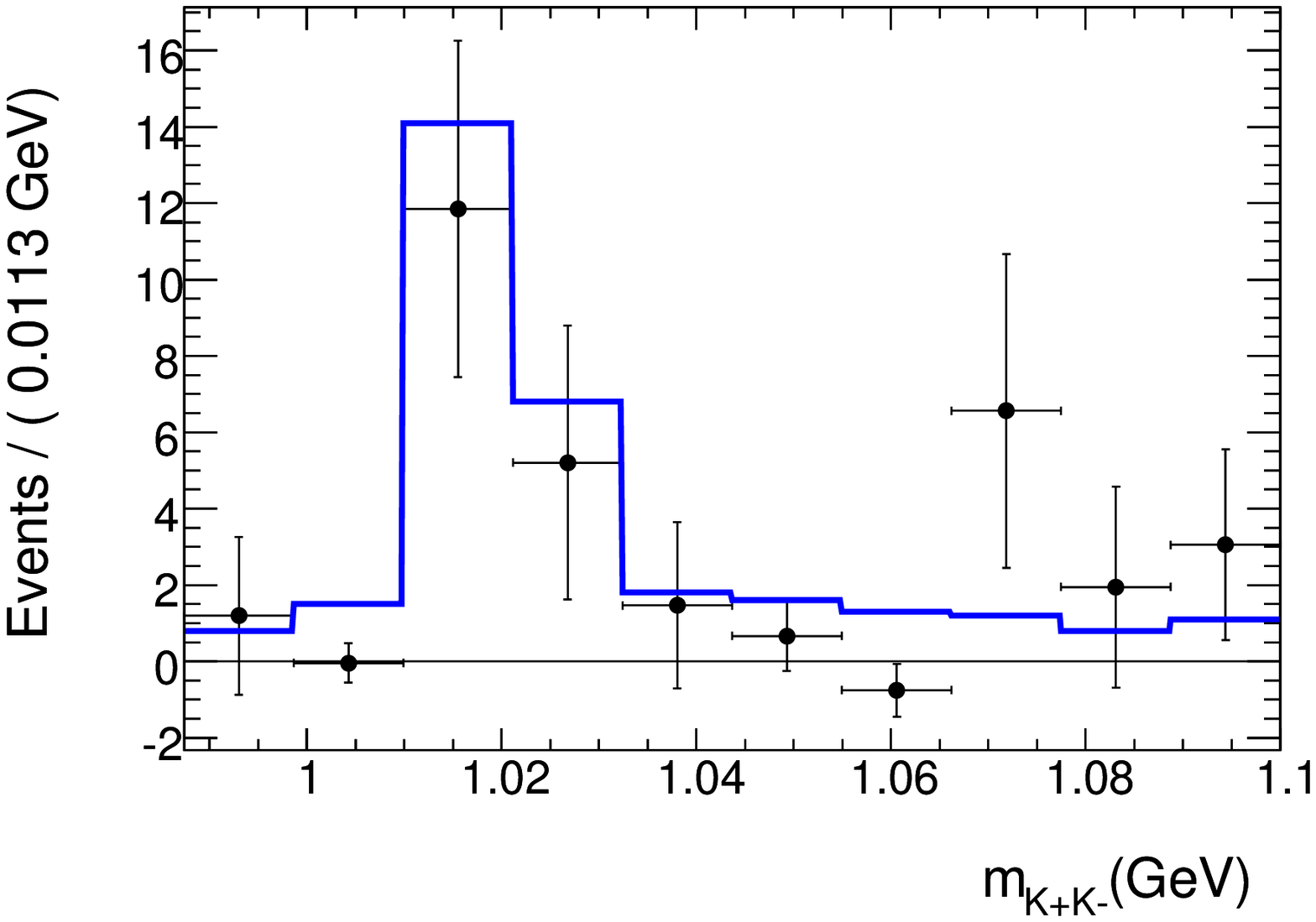} & \includegraphics[height=5.5cm]{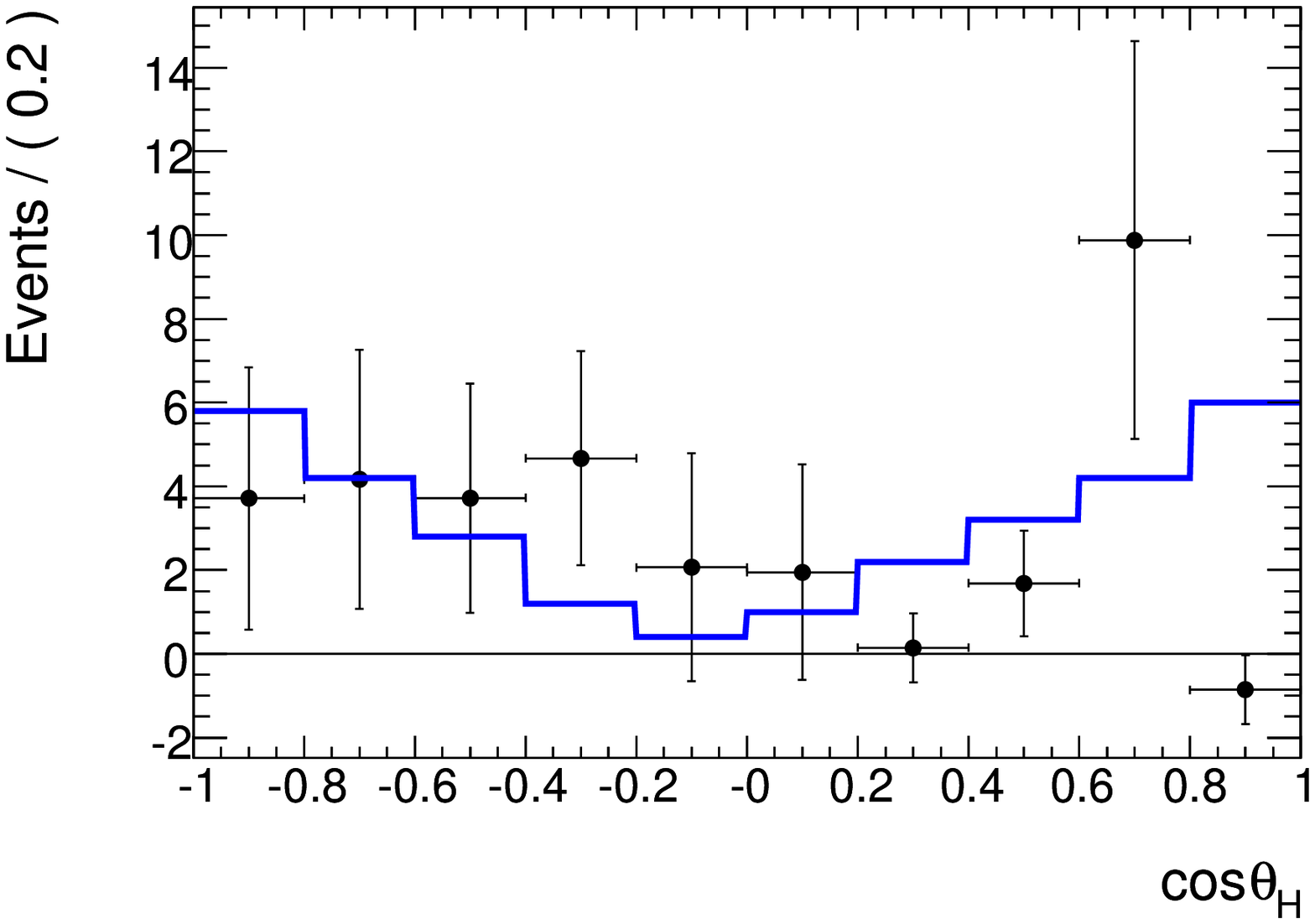} \\
c)\\
\includegraphics[height=5.5cm]{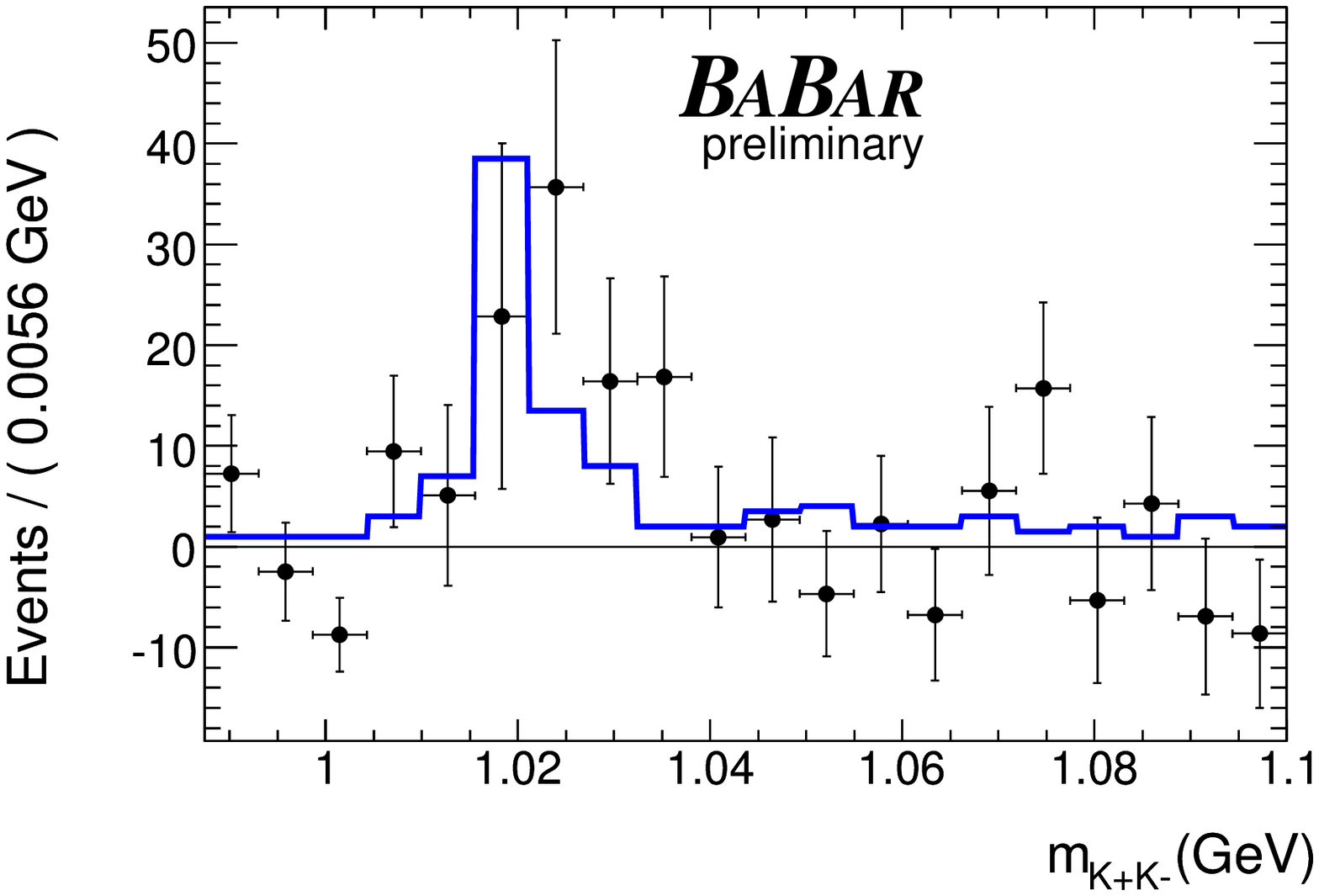} & \includegraphics[height=5.5cm]{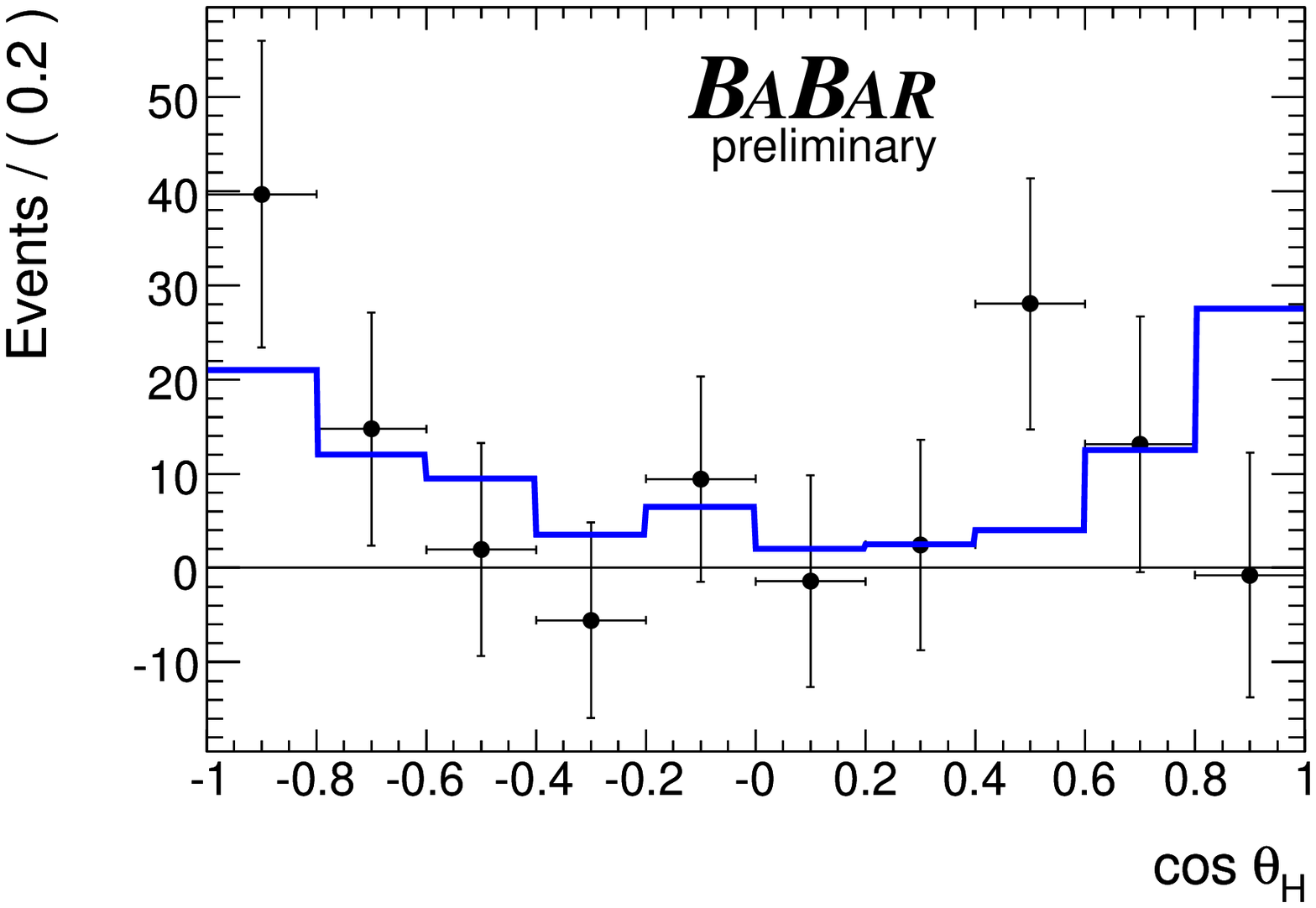} \\
\end{tabular}
\caption{For the low-$\Kp\Km$ mass Dalitz plot fit, distributions of the Dalitz plot variables (left) \mKK and (right) \cosH for signal events (points) 
compared with the fit PDF in the following sub-samples: a) \KKKspm, b) \KKKszz, c) \KKKl.}
\label{fg::dalitz-splot_low-mass-fit}
\end{figure}

\begin{figure}[ptb]
\center
\begin{tabular}{ll}
\includegraphics[width=8cm,height=4cm]{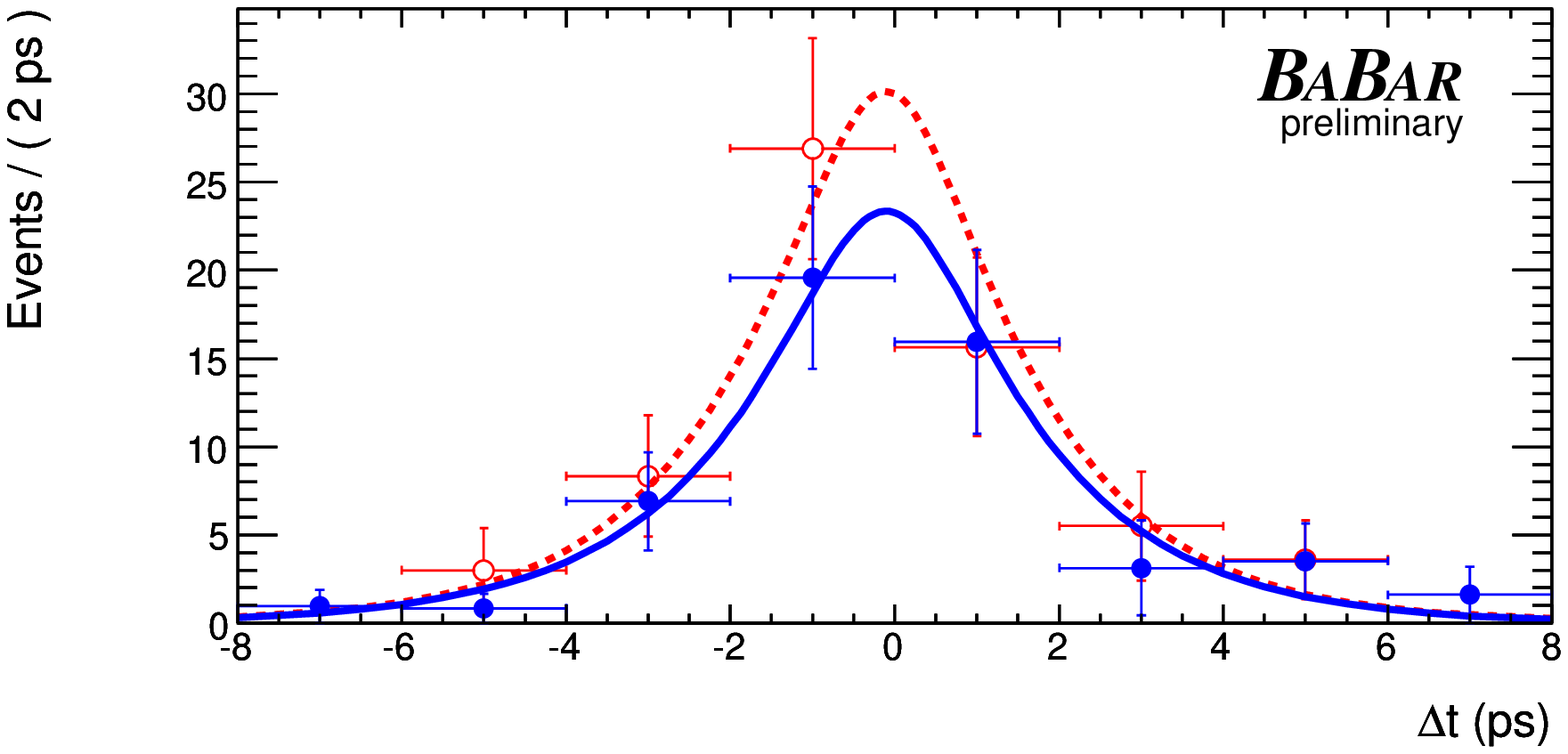}     &  \includegraphics[width=8cm,height=4cm]{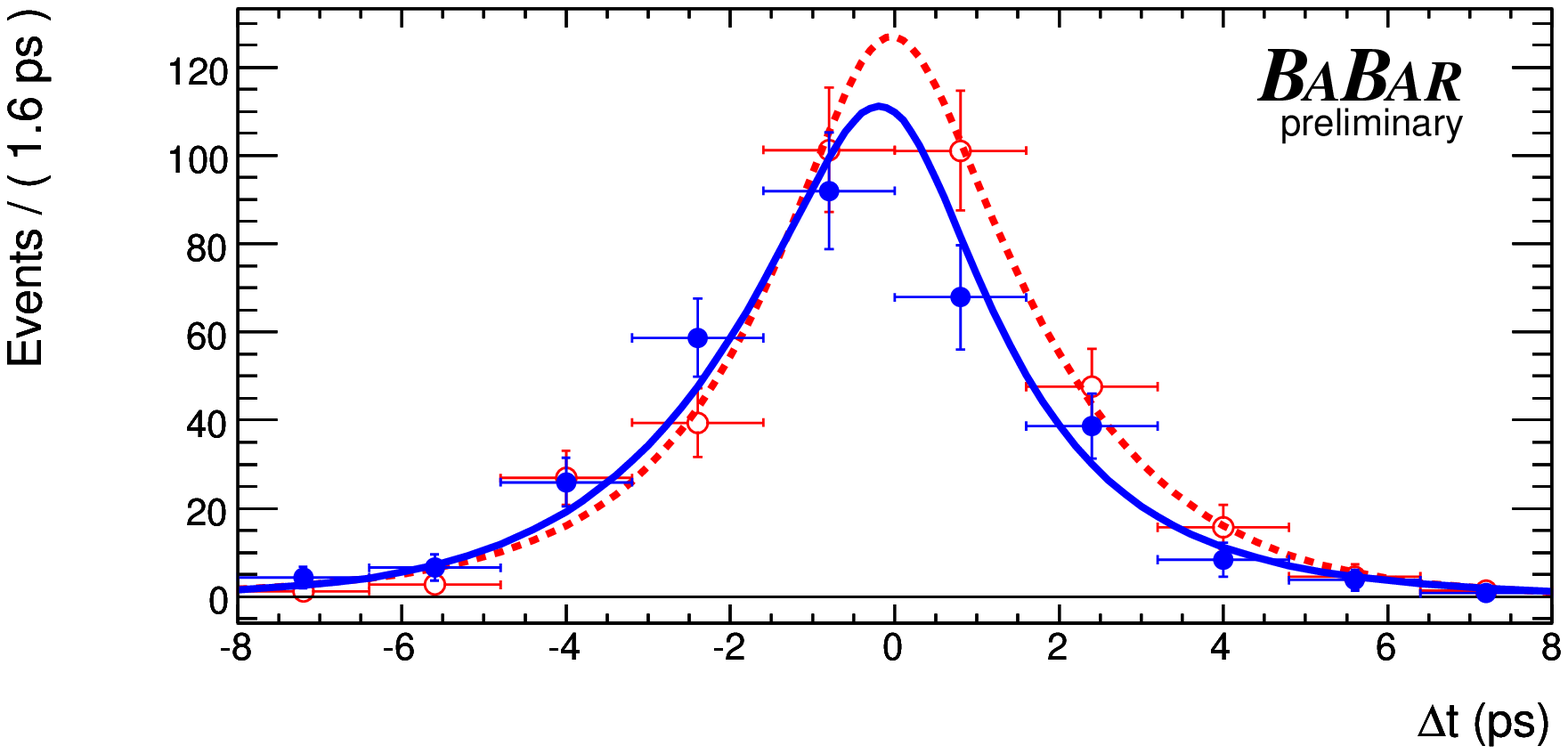}  \\
\includegraphics[width=8cm,height=4cm]{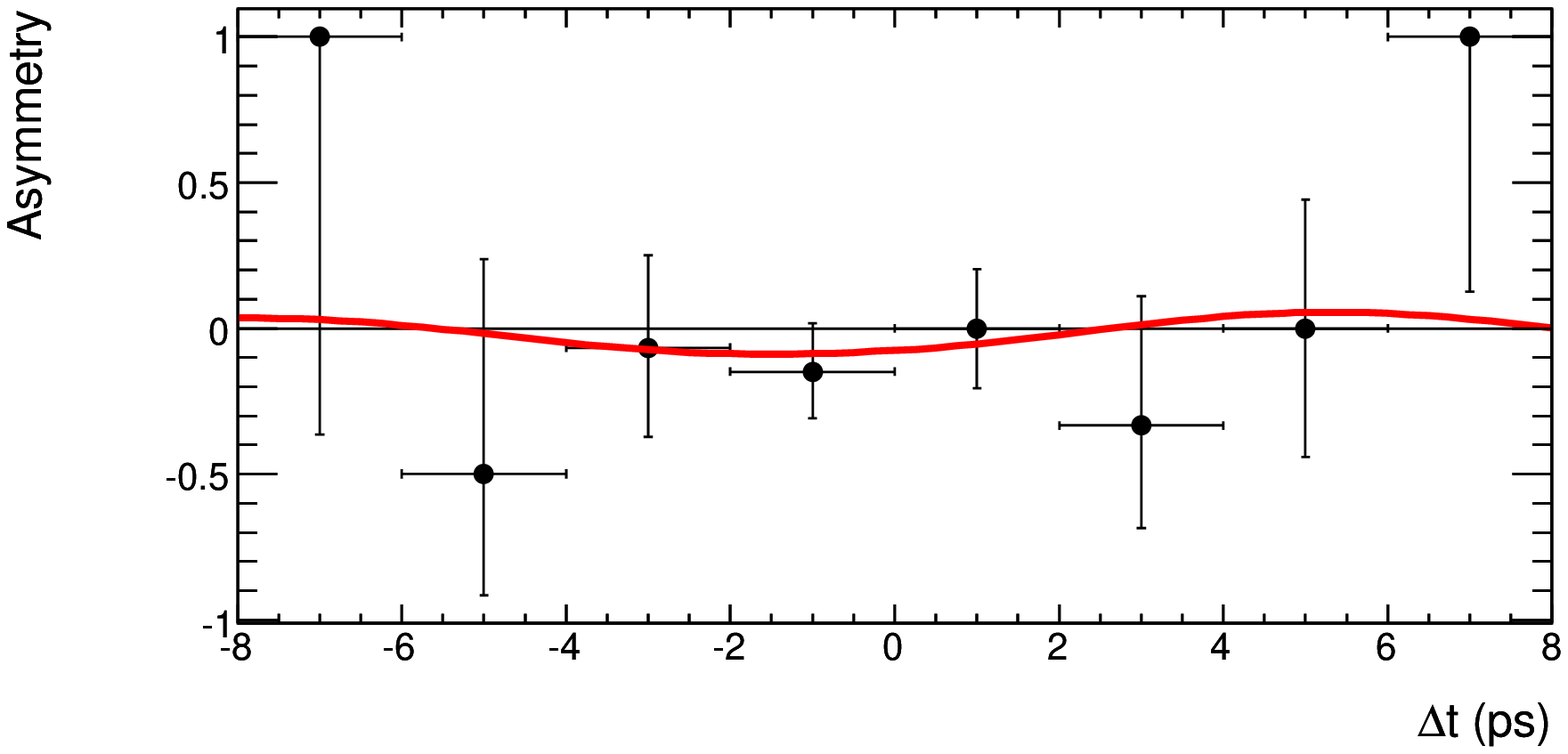} & \includegraphics[width=8cm,height=4cm]{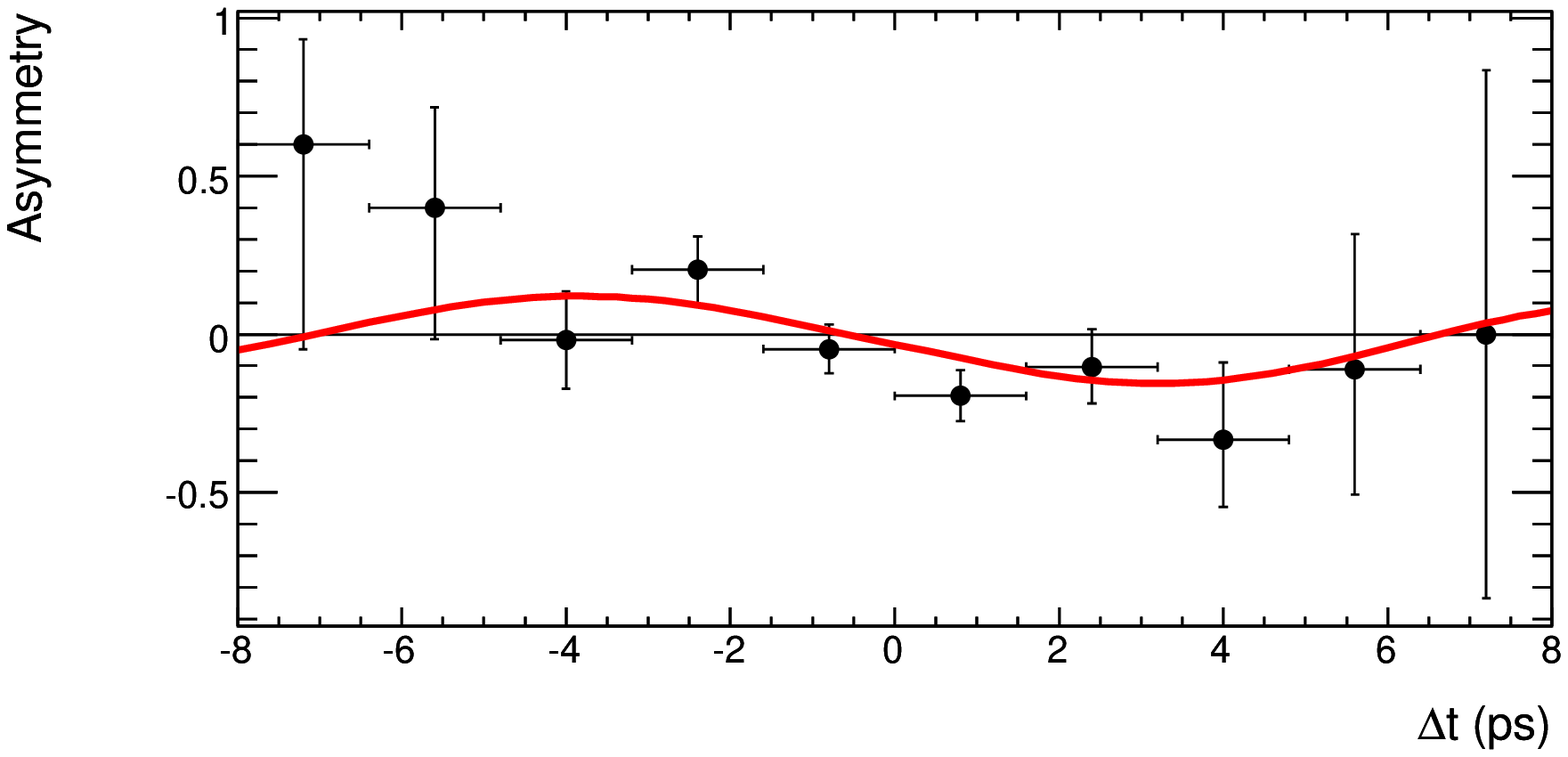}
\end{tabular}
\caption{(top) \deltat\ distributions and (bottom) asymmetries in the \KKKspm submode for (left) $1.0045 < \mKK < 1.0345~\gevcc$ and (right) the whole Dalitz plot. For the \deltat\ distributions, \Bz- (\Bzb-) tagged signal-weighted events are shown as filled (open) circles, with the PDF projection in solid blue (dashed red).}
\label{fg::dt-splot_low-mass-same-dcpv}
\end{figure}

In order to measure \CP-asymmetry parameters for components with
low-$\Kp\Km$ mass with reduced model-dependence from the rest of the
Dalitz plot, we select events using a cut of $\mKK <1.1~\gevcc$. 
After this selection, there are 836  \KKKspm candidates, 202 \KKKszz candidates, 
and 4923 \KKKl candidates remaining. 
The most significant contribution in this region comes from
$\phi(1020)\Kz$ and $f_0(980)\Kz$ decays, with a smaller contribution from
a low-$\Kp\Km$ tail of non-resonant decays.
We vary the isobar parameters for the $\phi(1020)$ and fix all other components to
the results of the full Dalitz plot fit.
We also vary the \CP amplitude and phase asymmetries for the $\phi(1020)$ and
$f_0(980)$. The asymmetry for the other components is fixed to the SM
expectation.  
We perform a fit to the \KKKspm submode,
then perform an additional fit to the entire \KKKz sample. We find signal yields of 
$252 \pm19$, $35 \pm 9$, and $195\pm33$ events for \KKKspm, \KKKszz, and \KKKl respectively. 
 Fig.~\ref{fg::dalitz-splot_low-mass-fit} shows projections of the
Dalitz plot distributions of events in this region.
The \CP-asymmetry results
are listed in Table~\ref{tab:low_mass_yields_cp}; the systematic
uncertainties will be described in
Sec.~\ref{sec:Systematics}. The left plots in Fig.~\ref{fg::dt-splot_low-mass-same-dcpv}
show distributions of $\Delta t$ for \Bz-tagged and \Bzb-tagged
events, and the asymmetry ${\cal A}(\Delta
t)=(N_{\Bz}-N_{\Bzb})/(N_{\Bz}+N_{\Bzb})$, obtained with the \sPlot
event-weighting technique~\cite{Pivk:2004ty}.  Correlation
coefficients $r$ between \CP\ parameters, found in the fit to the combined sample,
are also shown in Table~\ref{tab:low_mass_yields_cp}.

\begin{table}[h]
\center
\begin{tabular}{|l|rr|rrrr|}
\hline \hline
Name                            &       \multicolumn{2}{c|}{Fitted Value} 		& \multicolumn{4}{c|}{Correlation}	\\
                                & \KKKs			&               		& \multicolumn{4}{c|}{(Combined)}		\\
                                & ($\pip\pim$)      	&       Combined        	& 1	& 2	& 3 	& 4		\\
\hline
1 $\Acp(\phi\Kz)$               & $-0.10 \pm 0.23$      &  $-0.18 \pm 0.20 \pm 0.10 $   & 1.0	& -0.08 & -0.27	& 0.11		\\
2 $\betaeff (\phi\Kz)$          & $ 0.02 \pm 0.16$      &  $ 0.06 \pm 0.16 \pm 0.05 $   &	& 1.0	& 0.46	& 0.74		\\
3 $\Acp(f_0\Kz)$                & $ 0.36 \pm 0.33$      &  $ 0.45 \pm 0.28 \pm 0.10 $   &	&	& 1.0	& 0.20 		\\
4 $\betaeff (f_0\Kz)$           & $ 0.04 \pm 0.18$      &  $ 0.18 \pm 0.19 \pm 0.04 $   &	&	&	& 1.0		\\
\hline \hline
\end{tabular}
\caption{\CP-violation parameters for $\Bz\to\KKKz$ for $\mKK < 1.1~\gevcc$. For the combined \KKKz sample, the first error is statistical and the second is systematic. For the \KKKs submode, only statistical errors are shown.}
\label{tab:low_mass_yields_cp}
\end{table}

The decay $\Bz\to\phi\Kz$, with highly suppressed tree amplitudes, is,
in terms of theoretical uncertainty, the cleanest channel to interpret
possible deviations of the \CP-violation parameters from the SM
expectations. 
Values of \betaeff are consistent with the value found 
in $[c\bar{c}]\Kz$ decays~\cite{Aubert:2004zt,Abe:2005bt}.

As a consistency check we compare $\betaeff (\phi\Kz)$ against a
quasi-two-body approach (Q2B) that selects events around the $\phi$ resonance,
$1.0045 < \mKK < 1.0345~\gevcc$, in order to increase the purity
of $P$-wave decays.  We add separate \cosH and \mKK PDFs to
the likelihood to further suppress the $S$-wave decays.  From this fit
we find $166 \pm 15$ $\phi\KS$ and $151 \pm 22$ $\phi\KL$ signal
events. The \CP parameters in the channel $B^0\to\phi\KS$ only are:
$\spk = 0.10 \pm 0.29$ and $\cpk = 0.28 \pm 0.20$; in the channel
$B^0\to\phi\KL$: $\spk = 0.69 \pm 0.35$ and $\cpk = -0.28 \pm 0.33$,
with statistical errors only.  The simultaneous quasi-two-body fit to
both $\phi K^0$ and flavor decay modes yields the result $\spk = 0.39
\pm 0.23$ and $\cpk = 0.10 \pm 0.18$ which is fully consistent with
the Dalitz plot result.
In this comparison we neglect interference effects and
use the approximation  $\spk \approx \sin(2 \betaeff)(1-b^2)/(1+b^2)$ and $\cpk \approx -\Acp$
to relate the Dalitz plot \CP-violation parameters to the Q2B \CP-violation parameters for the $\phi\Kz$ decay.

We also measure the direct \CP\ asymmetry  in $\Bp \to \phi\Kp$ decays, defined as
$\Acp = (N_{B^-} - N_{B^+})/(N_{B^-} + N_{B^+})$.
In a fit for the $\phi\Kp$ and $\phi\Km$ yields we find ${\cal A}_{CP} = 0.046 \pm 0.046 \pm 0.017$.

\subsection{Average {\boldmath \CP} Asymmetry in {\boldmath $\Bz\to\KKKz$}}

We fit the average \CP-violation parameters $\betaeff ,~\Acp$
across the $\Bz\to\KKKz$ Dalitz plot, and remove an ambiguity
in the solution for the mixing angle $\betaeff \to
\pi/2- \betaeff$, present in previous measurements of
$\sin(2 \betaeff)$ in penguin decays.  In our analysis the
reflection is removed due to interference between $\CP$-even and
$\CP$-odd decays that give rise to a $\cos (2 \betaeff)$ term, in
addition to the $\sin (2 \betaeff)$ terms that come from the
interference of decays with and without mixing.  Fit results for \CP
parameters are listed in Table~\ref{tab:dalitz_average_cp}
for both the \KKKspm submode and the entire sample. \deltat\ projection
plots for signal-weighted events~\cite{Pivk:2004ty}, shown in the right plots of 
Fig.~\ref{fg::dt-splot_low-mass-same-dcpv}, clearly show a large phase
difference between \Bz and \Bzb\ decays. The correlation coefficients $r$
between the \CP\ parameters are given in Table~\ref{tab:dalitz_average_cp}.
The global correlation coefficients in the fit to the combined sample 
are $0.02$ and $0.08$ for $\betaeff$ and $\Acp$, respectively.

\begin{table}[h]
\center
\begin{tabular}{|l|rr|}
\hline \hline
Name                    &       \multicolumn{2}{c|}{Fitted Value} \\ 
                        &       \KKKspm         &       \multicolumn{1}{c|}{Combined} \\
\hline
$\Acp$        & $ -0.100 \pm 0.089$    &  $-0.034 \pm 0.079 \pm 0.025$  \\
\betaeff           & $ 0.354 \pm 0.083$    &  $0.361 \pm 0.079 \pm 0.037$ \\ 
$r$		& $ 0.003$		&       $0.013$ 	\\
\hline \hline
\end{tabular}
\caption{Average \CP-asymmetry parameters for the $\Bz\to\KKKz$ Dalitz plot. For the combined \KKKz sample, 
the first error is statistical and the second is systematic. For the \KKKs submode, only statistical errors are shown.}
\label{tab:dalitz_average_cp}
\end{table}

\begin{figure}[ptb]
\begin{center}
\begin{tabular}{c}
\includegraphics[height=6.5cm]{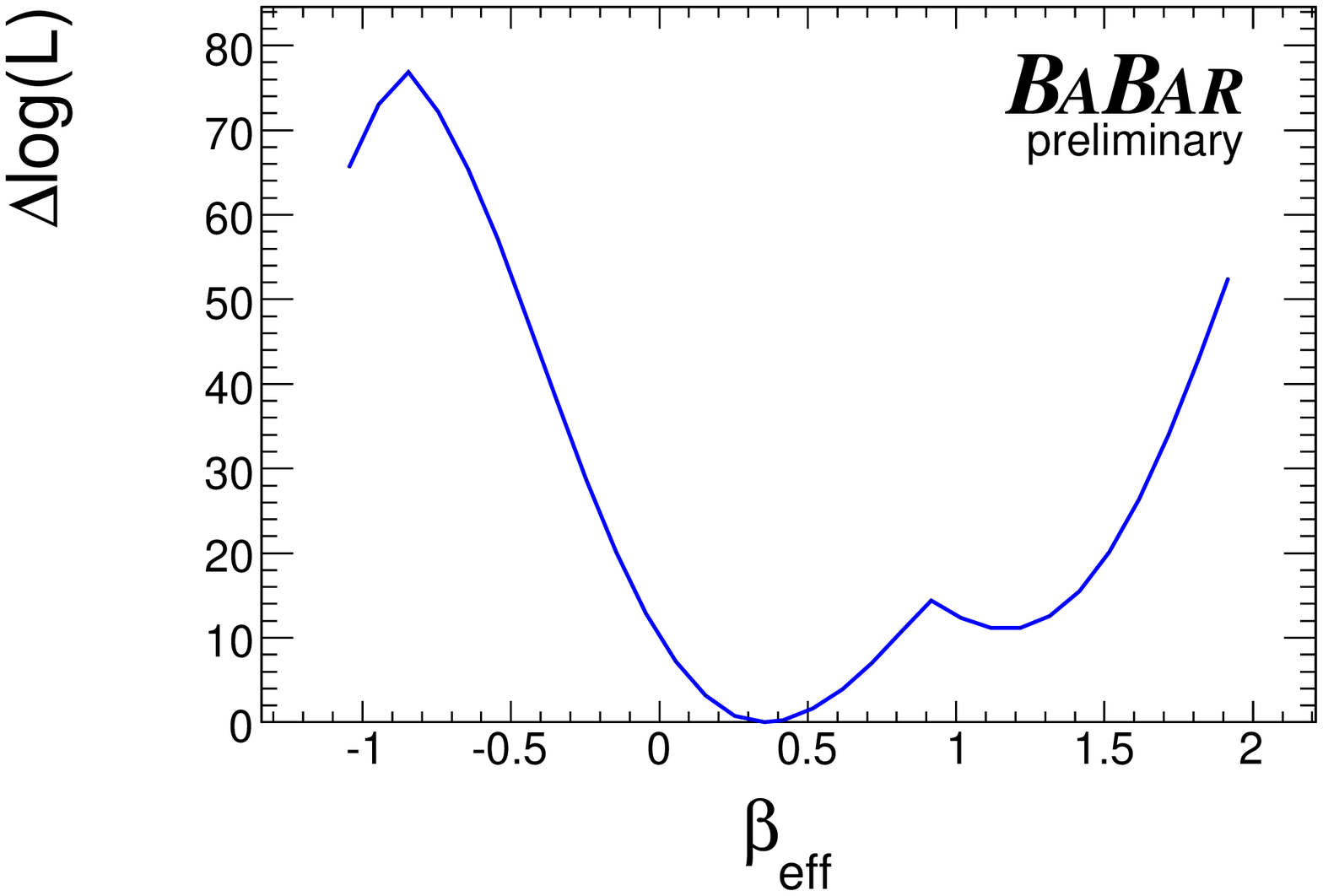} 
\vspace{-.5cm}
\end{tabular} 
\caption{\KKKspm: Change in the $\Delta\log({\cal L})$ value as a function of $\betaeff$.
\label{fig:kkkspm_betaeff_scan}}
\end{center}
\end{figure}

Using the \KKKspm subsample, we estimate the significance of the
nominal result for \betaeff compared
to the trigonometric reflection where $\betaeff \to
\pi/2- \betaeff$. In a collection of fits with both isobar
coefficients and \CP-asymmetry parameters allowed to vary, we
randomize the initial parameter values and evaluate the likelihood
separation between these two solutions. We find $\Delta\log({\cal
L})=10.4$, which excludes the reflection at a significance of 4.6 standard
deviations.  Note that a reflection $\betaeff \to \betaeff + \pi$
still remains since we measure the total phase difference between \Bz
and \Bzb decays ($2 \betaeff$). A scan of the change in likelihood as
a function of \betaeff is shown in Figure
\ref{fig:kkkspm_betaeff_scan}.

%
%

\section{SYSTEMATIC STUDIES}
\label{sec:Systematics}

We study systematic effects on the \CP-asymmetry parameters
due to fixed parameters in the event-selection (\mes and \DeltaE) PDFs.
We assign systematic errors  by comparing the fit with nominal parameters and 
with parameters smeared by their error, and assign the average difference as the systematic error.
We account for a potential fit bias using values observed in studies 
with MC samples generated with the nominal  Dalitz plot model.
We take the largest values of the bias observed in these studies as the systematic error.
We account for fixed \deltat\ resolution parameters, \Bz\ lifetime,
\Bz-\Bzb mixing and flavor tagging parameters. We also assign an error
due to interference between the CKM-suppressed $\bar{b}\to\bar{u} c\bar{d}$
and the favored $b\to c\bar{u}d$ amplitude for some tag-side $B$ decays~\cite{dcsd}.
Smaller errors due to beam-spot position uncertainty, detector alignment, and the
boost correction are based on studies done in charmonium decays.
In all fits we assume no direct \CP\ violation in decays dominated by the $b\to c$ transition ($\chi_{c0}\Kz$, $D_{(s)}K$).

\begin{table}[h]
\center
\begin{tabular}{|l|rrrr|rr|}
\hline \hline
Parameter &             \multicolumn{2}{c}{$\phi\Kz$} &         \multicolumn{2}{c|}{$f_0\Kz$}   &       \multicolumn{2}{c|}{$\KKKz$}    \\
                        & $\Acp$       &  \betaeff        &  $\Acp$    &  \betaeff &   $\Acp$   &  \betaeff        \\ \hline 
\hline
Event selection         &  0.00        &       0.01           &       0.00   &       0.00   & 0.003         &       0.002           \\
Fit Bias                &  0.04        &       0.01           &       0.04   &       0.02   & 0.004         &       0.010           \\
\deltat, vertexing      &  0.02        &       0.03           &       0.01   &       0.01   & 0.010         &       0.010           \\
Tagging                 &  0.01        &       0.00           &       0.01   &       0.00   & 0.021         &       0.002           \\
Dalitz model            &  0.09        &       0.03           &       0.09   &       0.03   & 0.011         &       0.035           \\
\hline
Total                   &  0.10        &       0.05           &       0.10   &       0.04   & 0.025         &       0.037           \\
\hline \hline
\end{tabular}
\caption{Summary of systematic errors on \CP-asymmetry parameters.  Errors for 
$\phi\Kz$ and $f_0\Kz$ \CP-parameters are based on the low-$\Kp\Km$-mass sample. 
The $\KKKz$ column refers to errors on average \CP\ parameters across the Dalitz plot.}
\label{tb::sys_errors}
\end{table}

In the  fit to low \Kp\Km masses, we extract \CP asymmetry parameters for  $\phi\Kz$ and $f_0\Kz$ decays
together with  isobar parameters for the $\phi\Kz$, while all other isobar parameters are fixed.
We evaluate the impact of the fixed parameters on the \CP\ violation results using samples of simulated events. 
We compare the reference fit result with the fit that has fixed Dalitz parameters smeared by their errors which 
are taken from Table~\ref{tab:isobars}.
The average difference in the \CP\ violation parameters is taken as the systematic error.
We also assign an error due to uncertainty in the resonant and non-resonant
line-shape parameters. For resonant components this includes the uncertainity in
the mass and width of the $X_0(1550)$. For that resonance, we replace our nominal parameters
with those found by different measurements: $m_r=1.491$~\gevcc, 
$\Gamma=0.145$~\gev~\cite{Garmash:2004wa}, and $m_r=1.507$~\gevcc, 
$\Gamma=0.109$~\gev~\cite{Eidelman:2004wy}. We take the largest observed difference
from the reference fit as the systematic error.
The non-resonant distributions are not motivated by theory so
we try several alternative non-resonant models which omit some of the dependences on $\Kp\KS$ and $\Km\KS$ masses (see Eq.~\ref{eq:nr}):
$e^{-\alpha m^2_{12} }$, 
$e^{-\alpha m^2_{12} } + c_{23} e^{i\phi_{23}} e^{-\alpha m^2_{23} }$,
and 
$e^{-\alpha m^2_{12} } + c_{13} e^{i\phi_{13}} e^{-\alpha m^2_{13} }$.
We also study the effect of the uncertainty of the shape parameter $\alpha$ on the \CP\ parameters.
The non-resonant events contribute to the background under the $\phi$ but their shape
is determined from the high-mass region. We therefore omit the non-resonant
terms and re-do the low-mass fits, and take the largest difference from the reference fit as a systematic error.
In the full Dalitz plot fit, the non-resonant term is the dominant contribution to the
sample and its omission is not reasonable. 
The mass resolution is neglected in the reference fit since it is small compared to the resonant width 
for all Dalitz plot components. We evaluate a potential bias on \CP-asymmetry parameters 
by repeating the fit with the Dalitz plot PDF convolved with the mass resolution function and take
the difference in the \CP\ parameters into the systematic error.

In the fit to the charge asymmetry  in $\Bp \to\phi\Kp$ we
consider systematic errors due to charge asymmetries in tracking and particle
identification (0.011), uncertainties in the parameterization of the signal 
Fisher PDF (0.003) and $B$ background content (0.012).  We add these 
contributions in quadrature to obtain the total systematic uncertainty 
on the direct \CP violation.

\section{CONCLUSIONS}
\label{sec:Summary}

In a fit to $\Bz \to \KKKspm$ decays, we analyze the
Dalitz plot distribution and measure the fractions to intermediate states, given in Table 1.
Subsequently, we extract \CP-asymmetry parameters from simultaneous fits to \KKKz final states with the neutral kaon
reconstructed as $\KS\to\pip\pim, \KS \to \piz\piz$, or $\KL$.
We further analyze the $\Kp\Km$ phase-space by computing moments of Legendre
polynomials in $\phi\Kp$ and $\KKKspm$ decays. 
We find the P-wave fraction to be $0.29\pm0.03$ averaged over the Dalitz plot, and
$0.89\pm0.01$ over the
$\phi(1020)$ resonance region ($1.0045 < \mKK < 1.0345$~\gevcc).  
 
From a fit to events at low $\Kp\Km$ masses, we find
$\betaeff=0.06\pm0.16\pm0.05$ for $\Bz\to\phi\Kz$ and
$0.18\pm0.19\pm0.04$ for $\Bz \to f_0 \Kz$, consistent with our
previous measurements~\cite{Aubert:2005ja} and with an update of the
previous method to the present dataset. We do not observe any
significant deviation of \CP parameters from the Standard Model values $\beta \simeq 0.37, \Acp = 0$.

In a fit to the full Dalitz plot, we find the CKM angle $\betaeff = 0.361 \pm 0.079 \pm 0.037$
to be compatible with the SM expectation.  Additionally, we resolve
the trigonometric ambiguity in the measurement of \betaeff at 4.6
standard deviations, which is the first such measurement in penguin
decays.

\section{ACKNOWLEDGMENTS}
\label{sec:Acknowledgments}

\input acknowledgements

\end{document}

%% file: authors_ICHEP2006.tex
\begin{center}
\small

The \babar\ Collaboration,
\bigskip

%
{B.~Aubert,}
{R.~Barate,}
{M.~Bona,}
{D.~Boutigny,}
{F.~Couderc,}
{Y.~Karyotakis,}
{J.~P.~Lees,}
{V.~Poireau,}
{V.~Tisserand,}
{A.~Zghiche}
\inst{Laboratoire de Physique des Particules, IN2P3/CNRS et Universit\'e de Savoie,
 F-74941 Annecy-Le-Vieux, France }
% ...

\end{center}\newpage

%% file: acknowledgements.tex
We are grateful for the 
extraordinary contributions of our \pep2\ colleagues in
achieving the excellent luminosity and machine conditions
that have made this work possible.
The success of this project also relies critically on the 
expertise and dedication of the computing organizations that 
support \babar.
The collaborating institutions wish to thank 
SLAC for its support and the kind hospitality extended to them. 
This work is supported by the
US Department of Energy
and National Science Foundation, the
Natural Sciences and Engineering Research Council (Canada),
Institute of High Energy Physics (China), the
Commissariat \`a l'Energie Atomique and
Institut National de Physique Nucl\'eaire et de Physique des Particules
(France), the
Bundesministerium f\"ur Bildung und Forschung and
Deutsche Forschungsgemeinschaft
(Germany), the
Istituto Nazionale di Fisica Nucleare (Italy),
the Foundation for Fundamental Research on Matter (The Netherlands),
the Research Council of Norway, the
Ministry of Science and Technology of the Russian Federation, 
Ministerio de Educaci\'on y Ciencia (Spain), and the
Particle Physics and Astronomy Research Council (United Kingdom). 
Individuals have received support from 
the Marie-Curie IEF program (European Union) and
the A. P. Sloan Foundation.

%% file: conf.bbl
\begin{thebibliography}{99}

\bibitem{conjugate}
 Unless explictly stated, charge conjugate is implied
 throughout this paper.


\bibitem{Eidelman:2004wy}
  S.~Eidelman {\it et al.}  [Particle Data Group],
  Phys.\ Lett.\ B {\bf 592}, 1 (2004).

\bibitem{Beneke:2005pu}
  M.~Beneke,
  Phys.\ Lett.\ B {\bf 620}, 143 (2005)
  [arXiv:hep-ph/0505075].



\bibitem{Buchalla:2005us}
  G.~Buchalla, G.~Hiller, Y.~Nir and G.~Raz,
  JHEP {\bf 0509}, 074 (2005)
  [arXiv:hep-ph/0503151].


\bibitem{Cheng:2005ug}
  H.~Y.~Cheng, C.~K.~Chua and A.~Soni,
  Phys.\ Rev.\ D {\bf 72}, 094003 (2005)
  [arXiv:hep-ph/0506268].

\bibitem{Aubert:2004zt}
  B.~Aubert {\it et al.}  [BABAR Collaboration],
  %
  Phys.\ Rev.\ Lett.\  {\bf 94}, 161803 (2005)
  [arXiv:hep-ex/0408127].

\bibitem{Abe:2005bt}
  K.~Abe {\it et al.}  [Belle Collaboration],
  %
  [arXiv:hep-ex/0507037].




\bibitem{Krokovny:2006sv}
  P.~Krokovny,
  [arXiv:hep-ex/0605023].

\bibitem{Aubert:2005ja}
  B.~Aubert {\it et al.}  [BABAR Collaboration],
  Phys.\ Rev.\ D {\bf 71}, 091102 (2005)
  [arXiv:hep-ex/0502019].


\bibitem{ref:babar}
  B.~Aubert {\it et al.}  [BABAR Collaboration],
  Nucl.\ Instrum.\ Meth.\ A {\bf 479}, 1 (2002)
  [arXiv:hep-ex/0105044].


\bibitem{Pivk:2004ty}
  M.~Pivk and F.~R.~Le Diberder,
  Nucl.\ Instrum.\ Meth.\ A {\bf 555}, 356 (2005)
  [arXiv:physics/0402083].

\bibitem{Roe:2004na}
  B.~P.~Roe, H.~J.~Yang, J.~Zhu, Y.~Liu, I.~Stancu and G.~McGregor,
  Nucl.\ Instrum.\ Meth.\ A {\bf 543}, 577 (2005)
  [arXiv:physics/0408124].

\bibitem{Yang:2005nz}
  H.~J.~Yang, B.~P.~Roe and J.~Zhu,
  Nucl.\ Instrum.\ Meth.\ A {\bf 555}, 370 (2005)
  [arXiv:physics/0508045].

\bibitem{Albrecht:1990cs}
  H.~Albrecht {\it et al.}  [ARGUS Collaboration],
  Z.\ Phys.\ C {\bf 48}, 543 (1990).

\bibitem{blatt}
	J.~M.~Blatt, V.~F.~Weisskopf,
	``Theoretical Nuclear Physics'',
	John Wiley \& Sons, New York (1952).

\bibitem{Zemach:1963bc}
C.~Zemach,
Phys.\ Rev.\  {\bf 133}, B1201 (1964).



\bibitem{Garmash:2004wa}
  A.~Garmash {\it et al.}  [BELLE Collaboration],
  Phys.\ Rev.\ D {\bf 71}, 092003 (2005)
  [arXiv:hep-ex/0412066].


\bibitem{Aubert:2006nu}
  B.~Aubert {\it et al.}  [BABAR Collaboration],
  [arXiv:hep-ex/0605003].

\bibitem{Aubert:2005kd}
  B.~Aubert {\it et al.}  [BABAR Collaboration],
  [arXiv:hep-ex/0507094].


\bibitem{Minkowski:2004xf}
  P.~Minkowski and W.~Ochs,
  Eur.\ Phys.\ J.\ C {\bf 39}, 71 (2005)
  [arXiv:hep-ph/0404194].


\bibitem{Ablikim:2004wn}
  M.~Ablikim {\it et al.}  [BES Collaboration],
  Phys.\ Lett.\ B {\bf 607}, 243 (2005)
  [arXiv:hep-ex/0411001].




\bibitem{Cheng:2002qu}
  H.~Y.~Cheng and K.~C.~Yang,
  Phys.\ Rev.\ D {\bf 66}, 054015 (2002)
  [arXiv:hep-ph/0205133].

\bibitem{Fajfer:2004cx}
  S.~Fajfer, T.~N.~Pham and A.~Prapotnik,
  Phys.\ Rev.\ D {\bf 70}, 034033 (2004)
  [arXiv:hep-ph/0405065].


\bibitem{Gronau:2005ax}
  M.~Gronau and J.~L.~Rosner,
  Phys.\ Rev.\ D {\bf 72}, 094031 (2005)
  [arXiv:hep-ph/0509155].


\bibitem{Aubert:2005ce}
  B.~Aubert {\it et al.}  [BABAR Collaboration],
  %
  Phys.\ Rev.\ D {\bf 72}, 072003 (2005)
  [arXiv:hep-ex/0507004].

\bibitem{dcsd}
O.~Long, M.~Baak, R.~Cahn, and D.~Kirkby, \jprd{68}, 034010 (2003).

\end{thebibliography}
